\documentclass[letterpaper,11pt]{article}

\usepackage{jheppub}

\usepackage{multirow}

\usepackage{subfig}
\usepackage{xspace}
\usepackage{xcolor}
\usepackage[countmax]{subfloat}
\usepackage{amsthm}

\usepackage{amsmath}

\allowdisplaybreaks

\usepackage{color}
\definecolor{darkblue}{rgb}{0,0,0.5}

\DeclareRobustCommand{\Sec}[1]{Sec.~\ref{#1}}

\DeclareRobustCommand{\App}[1]{App.~\ref{#1}}

\DeclareRobustCommand{\Fig}[1]{Fig.~\ref{#1}}

\DeclareRobustCommand{\Eq}[1]{Eq.~(\ref{#1})}
\DeclareRobustCommand{\Eqs}[2]{Eqs.~(\ref{#1}) and (\ref{#2})}
\DeclareRobustCommand{\Refnew}[1]{Ref.~\cite{#1}}
\DeclareRobustCommand{\Refs}[1]{Refs.~\cite{#1}}

\title{
Binary Discrimination Through Next-to-Leading Order
}

\author{Andrew J. Larkoski}

\affiliation{Department of Physics and Astronomy, University of California, Los Angeles, CA 90095, USA\\
Mani L. Bhaumik Institute for Theoretical Physics, University of California, Los Angeles, CA 90095, USA}
\emailAdd{larkoski@ucla.edu}

\abstract{
Binary discrimination between well-defined signal and background datasets is a problem of fundamental importance in particle physics.  With detailed event simulation and the advent of extensive deep learning tools, identification of the likelihood ratio has typically been reserved as a computational problem.  However, this approach can obscure overtraining or excessive sensitivity to tuned features of the simulation that may not be well-defined theoretically. Here, we present the first analysis of binary discrimination for signal and background distributions for which their likelihood ratio is infrared and collinear safe, and can therefore be calculated order-by-order in perturbation theory.  We present explicit, general formulas for receiver operator characteristic curves and the area under it through next-to-leading order.  As a demonstration of this formalism, we apply it to discrimination of highly-boosted Higgs decays from gluon splitting to bottom quarks.  Effects at next-to-leading order are first sensitive to the flow of color in the jet and significantly modify discrimination performance at leading-order.  In the limit of infinite boost, these events can be perfectly discriminated because only the gluon will radiate at finite angles from the bottom quarks, and we find that large effects persist at energies accessible at the Large Hadron Collider.  Next-to-leading order is therefore required to qualitatively understand results using machine-learning methods. 
}

\begin{document}
\maketitle

\section{Introduction}

Given an ensemble of events from a particle collider experiment, such as the Large Hadron Collider, a natural first question is to what process or processes each individual event corresponds.  Quantum mechanically, this question cannot be answered with certainty, of course, but can be at a statistical level, with each event having some probability to have been a manifestation of one of the many possible processes.  Reduced to a binary discrimination problem, where all we have to classify is the probability of an event to be of signal or background process type, it is well-known that the optimal discriminant observable is (monotonic in) the likelihood ratio, the ratio of probability distributions of background to signal, by the Neyman-Pearson lemma \cite{Neyman:1933wgr}.  Further, given the availability of high-fidelity data simulators \cite{Bierlich:2022pfr,Bellm:2019zci,Sherpa:2019gpd} and the enormous dimensionality of particle physics events, this classification problem is naturally relegated to machine learning, whose interest in particle physics has exploded recently.  See \Refs{Larkoski:2017jix,Guest:2018yhq,Albertsson:2018maf,Radovic:2018dip,Carleo:2019ptp,Bourilkov:2019yoi,Schwartz:2021ftp,Karagiorgi:2021ngt,Boehnlein:2021eym,Shanahan:2022ifi} for reviews of the utility of machine learning in particle physics.

Nearly uniquely in fields that employ machine learning for data analysis, high-energy particle physics has a precise underlying theoretical description from which calculations can be performed.  Machine learning studies that exclusively employ simulated data therefore provide only an incomplete picture that can significantly obscure the features that are being learned to establish the likelihood for a particular problem of interest.  Further, because event simulators include numerous models and parameters on top of a perturbative parton shower or fixed-order matrix elements, it can be unclear what the machine is learning and if the features are robust physics, or simply idiosyncrasies of the simulation.  Therefore, in parallel with machine learning that pushes analysis boundaries, one would also desire predictions for discrimination performance that is well-defined theoretically, calculable from first-principles, and systematically-improvable.  For classification of events or individual jets, collimated streams of high-energy particles, at colliders, this requires that the likelihood ratio is infrared and collinear (IRC) safe  \cite{Sterman:1977wj,Ellis:1996mzs} and therefore can be predicted within the perturbation theory of quantum chromodynamics (QCD).

However, for many classification problems it can be argued that the likelihood is not IRC safe \cite{Soyez:2012hv}, or IRC safety is only manifest once all-orders effects are included \cite{Larkoski:2019nwj,Kasieczka:2020nyd}, and so a perturbative analysis can be impossible or at the very least obscured.  However, through the exercise of constructing an IRC safe likelihood ratio, theoretically-improved definitions of ``signal'' and ``background'' may present themselves and correspondingly render what was once deemed impossible now possible.  Here, we present the first systematic analysis of IRC safe likelihoods for binary classification tasks order-by-order in perturbation theory.  Because the likelihood ratio changes order-by-order because the distribution of radiation in jets or events changes order-by-order, there is a feedback and crosstalk between perturbative expansions of the distribution of the likelihood ratio and the definition of the likelihood ratio itself.  This results in an intricate perturbative expansion, but the necessary properties of the likelihood ratio to be IRC safe become obvious: the likelihood must be well-defined at lowest, leading order, and further one must map the phase space of emissions at higher orders down to that of leading order (LO) in an IRC safe way.

In this paper, we present results through next-to-leading order (NLO) in the strong coupling $\alpha_s$ for a generic likelihood ratio, signal and background distributions of the likelihood, and its receiver operating characteristic (ROC) curve that quantifies discrimination power.  As a concrete application of this framework, we apply it to the problem of discrimination of Higgs boson decays to bottom quarks, $H\to b\bar b$, from massive fragmentation of gluons to bottom quarks, $g\to b\bar b$, in the highly-boosted limit.  We find that next-to-leading order is crucial for qualitative understanding of machine learning results, because that is the first order sensitive to the flow of color, and therefore the radiation pattern of emissions, off of the final state bottom quarks.  In the process, we discuss the necessity of IRC safe NLO-to-LO phase space maps, robust and IRC safe flavor tagging of the bottom quarks, and practicalities of numerical integration of infrared divergent matrix elements.  This correspondingly provides a foundation for future analyses to establish theoretical predictions for various other discrimination problems.

This paper is organized as follows.  In \Sec{sec:rocnlo}, we discuss the general framework of binary discrimination through next-to-leading order and present formulas that apply to any such problem that admits an IRC safe likelihood ratio.  In \Sec{sec:hbb}, we apply the general results to $H\to b\bar b$ from $g\to b\bar b$ discrimination, establishing baseline performance at leading-order, and then improving the description at next-to-leading order.  We work in the highly-boosted limit, assuming that all final state particles are contained within collimated jets.  As a quantifiable metric for discrimination performance, we show that the area under the ROC curve (AUC) decreases by about a factor of 3 in going from leading- to next-to-leading order, demonstrating the importance of accounting for additional radiation in the jets.  We conclude in \Sec{sec:conc}, and just touch on continuing to next-to-next-to-leading order for improved uncertainty analysis and for re-evaluating classification tasks like hadronic top quark decay from this minimal and IRC safe perspective.  Appendices collect results necessary for numerical integration of next-to-leading order matrix elements.

\section{The ROC Curve Through NLO}\label{sec:rocnlo}

In this section, we present the derivation of the ROC curve for optimal binary discrimination through next-to-leading order in the strong coupling, $\alpha_s$.  We assume that we have two samples of events, signal $S$ and background $B$, and their respective particle momentum distributions are measured on phase space $\Phi$.  An expansion in $\alpha_s$ requires the property of infrared and collinear safety, and so particle momentum is the only information that is accessible to our discriminant.  The signal and background probability distributions on phase space can be expanded in powers of $\alpha_s$ as
\begin{align}
p_S(\Phi)& = p_S^{(0)}(\Phi^{(0)})+\frac{\alpha_s}{2\pi}\,p_S^{(1)}(\Phi^{(1)})+{\cal O}(\alpha_s^2)\,,\\
p_B(\Phi)& = p_B^{(0)}(\Phi^{(0)})+\frac{\alpha_s}{2\pi}\,p_B^{(1)}(\Phi^{(1)})+{\cal O}(\alpha_s^2)\,,
\end{align}
where the superscript denotes the order of that term in $\alpha_s$ and the corresponding phase space for the number of particles produced at that order. As probability distributions, they must be normalized to integrate to unity, which requires dividing by their total integral over all of phase space.  Starting from normalization of the leading-order distribution 
\begin{align}
1=\int d\Phi^{(0)}\, p^{(0)}(\Phi^{(0)})\,,
\end{align}
our normalization prescription for the contributions to the distribution at higher-orders is defined via the expansion of the ratio
\begin{align}\label{eq:normdef}
\frac{p^{(0)}(\Phi^{(0)})+\frac{\alpha_s}{2\pi}\,p^{(1)}(\Phi^{(1)})+\cdots}{1+\frac{\alpha_s}{2\pi}\int d\Phi^{(1)}\,p^{(1)}(\Phi^{(1)})+\cdots} &=p^{(0)}(\Phi^{(0)})\\
&\hspace{1cm}+\frac{\alpha_s}{2\pi}\,p^{(1)}(\Phi^{(1)}) - \frac{\alpha_s}{2\pi}\,p^{(0)}(\Phi^{(0)})\, \int d\Phi^{(1)}\,p^{(1)}(\Phi^{(1)})+\cdots\nonumber\,,
\end{align}
where terms at higher orders in $\alpha_s$ are suppressed in the ellipses.  At order $\alpha_s$, there are two contributions whose sum integrates to 0:
\begin{align}
\frac{\alpha_s}{2\pi}\int d\Phi^{(1)}\,p^{(1)}(\Phi^{(1)}) - \frac{\alpha_s}{2\pi}\int d\Phi^{(0)}\,p^{(0)}(\Phi^{(0)})\, \int d\Phi^{(1)}\,p^{(1)}(\Phi^{(1)}) = 0\,.
\end{align}
Because it lives in a higher-dimensional phase space, we refer to $p^{(1)}(\Phi^{(1)})$ as the real contribution and thus
\begin{align}
-p^{(0)}(\Phi^{(0)})\, \int d\Phi^{(1)}\,p^{(1)}(\Phi^{(1)}) \equiv p^{(0)}(\Phi^{(0)})\, {\cal V}^{(1)}\,,
\end{align}
as the virtual contribution.  Real, virtual, and mixed real-virtual contributions at higher orders are defined similarly, from simply Taylor expansion of \Eq{eq:normdef} to higher orders in $\alpha_s$.

By the Neyman-Pearson lemma \cite{Neyman:1933wgr}, a function monotonic in the likelihood ratio is the optimal binary discrimination observable.  The likelihood ratio ${\cal L}$ defined as a function on phase space is the ratio of the signal and background distributions
\begin{align}
\hat {\cal L}(\Phi) \equiv \frac{p_B(\Phi)}{p_S(\Phi)}\,,
\end{align}
so that ${\cal L}\to 0$ is the signal-rich region of phase space and pure background is pushed to ${\cal L}\to\infty$.  Here, we use the caret notation $\hat {\cal L}(\Phi)$ to denote the likelihood as a particular function of phase space $\Phi$, while ${\cal L}$ is its value.  Because it is defined through perturbative distributions itself, the likelihood ratio has a perturbative expansion, where
\begin{align}
\hat {\cal L}(\Phi) &= \frac{p_B^{(0)}(\Phi^{(0)})+\frac{\alpha_s}{2\pi}\,p_B^{(1)}(\Phi^{(1)})+\frac{\alpha_s}{2\pi}{\cal V}_B^{(1)}\,p_B^{(0)}(\Phi^{(0)})+{\cal O}(\alpha_s^2)}{p_S^{(0)}(\Phi^{(0)})+\frac{\alpha_s}{2\pi}\,p_S^{(1)}(\Phi^{(1)})+\frac{\alpha_s}{2\pi}{\cal V}_S^{(1)}\,p_S^{(0)}(\Phi^{(0)})+{\cal O}(\alpha_s^2)}\\
&=\frac{p_B^{(0)}(\Phi^{(0)})}{p_S^{(0)}(\Phi^{(0)})}+\frac{\alpha_s}{2\pi}\, \frac{p_B^{(0)}(\Phi^{(0)})}{p_S^{(0)}(\Phi^{(0)})}\left(
\frac{p_B^{(1)}(\Phi^{(1)})}{p_B^{(0)}(\Phi^{(0)})}+{\cal V}^{(1)}_B-\frac{p_S^{(1)}(\Phi^{(1)})}{p_S^{(0)}(\Phi^{(0)})}-{\cal V}^{(1)}_S
\right)+{\cal O}(\alpha_s^2)
\nonumber\\
&=\hat {\cal L}^{(0)}(\Phi)+\frac{\alpha_s}{2\pi}\, \hat {\cal L}^{(1)}(\Phi)+{\cal O}(\alpha_s^2)
\nonumber\,,
\end{align}
where we have introduced notation for the likelihood at order-$n$ in the perturbative expansion.  Note that this perturbative expansion of the likelihood ratio only makes sense if it is IRC safe.  Specifically, the leading-order likelihood ratio $\hat {\cal L}^{(0)}(\Phi)$ must be IRC safe, as it is a ratio of distributions and appears as a factor at every order in the expansion.

\subsection{Perturbative Expansion of the Cumulative Distribution of the Likelihood}\label{sec:pertcumlike}

With these perturbative expansions of the signal and background distributions and their likelihood ratio, we can work to calculate the distribution of the likelihood on the respective event samples.  However, here we will forgo that analysis and jump directly to calculation of the receiver operating characteristic or ROC curve.  The ROC curve is a signal versus background efficiency curve that quantifies the true and false positive rate for a sliding cut on the likelihood.  It can be defined through the signal and background cumulative distributions of the likelihood ratio as
\begin{align}
\text{ROC}(x) = \Sigma_B\left(
\Sigma_S^{-1}(x)
\right)\,,
\end{align}
where $\Sigma_B({\cal L})$ is the cumulative distribution of the likelihood on the background sample and $\Sigma_S^{-1}(x)$ is the inverse cumulative distribution of the likelihood on the signal sample, evaluated at quantile $x\in[0,1]$.  With the expansions established above, we will determine the expansion of the ROC curve through next-to-leading order in the coupling $\alpha_s$.

The first step is to determine the expansion of the cumulative distribution of the likelihood ratio, $\Sigma({\cal L})$.  It is defined as the total probability for the likelihood to be less than a given value, where
\begin{align}
\Sigma({\cal L})\equiv \int d\Phi\, p(\Phi)\,\Theta\left({\cal L}-\hat{\cal L}(\Phi)\right)\,.
\end{align}
To determine its expansion in $\alpha_s$, we expand both the probability $p(\Phi)$ and the likelihood $\hat{\cal L}(\Phi)$ on phase space, where
\begin{align}\label{eq:likecumdist}
\Sigma({\cal L})&= \int d\Phi\, \left(p^{(0)}(\Phi^{(0)})+\frac{\alpha_s}{2\pi}\, p^{(1)}(\Phi^{(1)})+\frac{\alpha_s}{2\pi}\,{\cal V}^{(1)}\, p^{(0)}(\Phi^{(0)})+\cdots\right)\\
&
\hspace{6cm}\times\Theta\left({\cal L}-\hat{\cal L}^{(0)}(\Phi)-\frac{\alpha_s}{2\pi}\, \hat{\cal L}^{(1)}(\Phi)+\cdots\right)\nonumber\\
&=\int d\Phi^{(0)}\, p^{(0)}(\Phi^{(0)})\, \Theta\left(
{\cal L}-\hat {\cal L}^{(0)}(\Phi^{(0)})
\right)+\frac{\alpha_s}{2\pi} \int d\Phi^{(1)}\, p^{(1)}(\Phi^{(1)})\,\Theta\left({\cal L}-\hat{\cal L}^{(0)}(\Phi^{(0)})\right)\nonumber\\
&\hspace{1cm}+\frac{\alpha_s}{2\pi}\,{\cal V}^{(1)} \int d\Phi^{(0)}\, p^{(0)}(\Phi^{(0)})\,\Theta\left({\cal L}-\hat{\cal L}^{(0)}(\Phi^{(0)})\right)\nonumber\\
&
\hspace{1cm}-\frac{\alpha_s}{2\pi}\int d\Phi\, p^{(0)}(\Phi^{(0)})\, \hat{\cal L}^{(1)}(\Phi)\, \delta\left(
{\cal L}-\hat{\cal L}^{(0)}(\Phi^{(0)})
\right)+{\cal O}(\alpha_s^2)\,.\nonumber
\end{align}
The final line, containing dependence on the likelihood at next-to-leading order, implicitly mixes contributions that live in different phase spaces, so it is useful to expand that out to identify individual contributions.  Note that
\begin{align}
\hat {\cal L}^{(1)}(\Phi) = \hat {\cal L}^{(0)}(\Phi^{(0)})\, \left(
\frac{p_B^{(1)}(\Phi^{(1)})}{p_B^{(0)}(\Phi^{(0)})}+{\cal V}^{(1)}_B-\frac{p_S^{(1)}(\Phi^{(1)})}{p_S^{(0)}(\Phi^{(0)})}-{\cal V}^{(1)}_S
\right)\,.
\end{align}
Thus, the integral on the final line of \Eq{eq:likecumdist} can be expressed as
\begin{align}
&\int d\Phi\, p^{(0)}(\Phi^{(0)})\, \hat{\cal L}^{(1)}(\Phi)\, \delta\left(
{\cal L}-\hat{\cal L}^{(0)}(\Phi^{(0)})
\right) \\
&\hspace{1cm}
={\cal L}\left(
{\cal V}^{(1)}_B-{\cal V}^{(1)}_S
\right) \int d\Phi^{(0)}\, p^{(0)}(\Phi^{(0)})\, \, \delta\left(
{\cal L}-\hat{\cal L}^{(0)}(\Phi^{(0)})
\right)\nonumber\\
&
\hspace{2cm}+{\cal L}\int d\Phi^{(1)}\, p^{(0)}(\Phi^{(0)})\left(
\frac{p_B^{(1)}(\Phi^{(1)})}{p_B^{(0)}(\Phi^{(0)})}-\frac{p_S^{(1)}(\Phi^{(1)})}{p_S^{(0)}(\Phi^{(0)})}
\right)\delta\left(
{\cal L}-\hat{\cal L}^{(0)}(\Phi^{(0)})
\right)\nonumber\,.
\end{align}

We can now put these results together to construct the signal and background cumulative distributions of the likelihood through next-to-leading order.  For the background, we have
\begin{align}
\Sigma_B({\cal L})&=\left(1+\frac{\alpha_s}{2\pi}\,{\cal V}_B^{(1)}\right)\int d\Phi^{(0)}\, p_B^{(0)}(\Phi^{(0)})\, \Theta\left(
{\cal L}-\hat {\cal L}^{(0)}(\Phi^{(0)})
\right)\\
&
\hspace{1cm}+\frac{\alpha_s}{2\pi} \int d\Phi^{(1)}\, p_B^{(1)}(\Phi^{(1)})\,\Theta\left({\cal L}-\hat{\cal L}^{(0)}(\Phi^{(0)})\right)\nonumber\\
&
\hspace{1cm}-\frac{\alpha_s}{2\pi}\,{\cal L}\left(
{\cal V}^{(1)}_B-{\cal V}^{(1)}_S
\right) \int d\Phi^{(0)}\, p_B^{(0)}(\Phi^{(0)})\, \, \delta\left(
{\cal L}-\hat{\cal L}^{(0)}(\Phi^{(0)})
\right)\nonumber\\
&
\hspace{1cm}-\frac{\alpha_s}{2\pi}\,{\cal L}\int d\Phi^{(1)} \left(
p_B^{(1)}(\Phi^{(1)})-{\cal L}\,p_S^{(1)}(\Phi^{(1)})
\right)\delta\left(
{\cal L}-\hat{\cal L}^{(0)}(\Phi^{(0)})
\right)\nonumber+{\cal O}(\alpha_s^2)\,.
\end{align}
The signal distribution is
\begin{align}
\Sigma_S({\cal L})&=\left(1+\frac{\alpha_s}{2\pi}\,{\cal V}_S^{(1)}\right)\int d\Phi^{(0)}\, p_S^{(0)}(\Phi^{(0)})\, \Theta\left(
{\cal L}-\hat {\cal L}^{(0)}(\Phi^{(0)})
\right)\\
&
\hspace{1cm}+\frac{\alpha_s}{2\pi} \int d\Phi^{(1)}\, p_S^{(1)}(\Phi^{(1)})\,\Theta\left({\cal L}-\hat{\cal L}^{(0)}(\Phi^{(0)})\right)\nonumber\\
&
\hspace{1cm}-\frac{\alpha_s}{2\pi}\,{\cal L}\left(
{\cal V}^{(1)}_B-{\cal V}^{(1)}_S
\right) \int d\Phi^{(0)}\, p_S^{(0)}(\Phi^{(0)})\, \, \delta\left(
{\cal L}-\hat{\cal L}^{(0)}(\Phi^{(0)})
\right)\nonumber\\
&
\hspace{1cm}-\frac{\alpha_s}{2\pi}\int d\Phi^{(1)} \left(
p_B^{(1)}(\Phi^{(1)})-{\cal L}\,p_S^{(1)}(\Phi^{(1)})
\right)\delta\left(
{\cal L}-\hat{\cal L}^{(0)}(\Phi^{(0)})
\right)\nonumber+{\cal O}(\alpha_s^2)\,.
\end{align}
Both of these distributions have an expansion in $\alpha_s$, and we denote it in the same way as the underlying distributions on phase space, where\begin{align}
\Sigma({\cal L})&=\Sigma^{(0)}({\cal L})+\frac{\alpha_s}{2\pi}\,\Sigma^{(1)}({\cal L})+{\cal O}(\alpha_s^2)\,.
\end{align}

This form explicitly demonstrates that we must apply a procedure to map next-to-leading order phase space $d\Phi^{(1)}$ onto leading order phase space $d\Phi^{(0)}$, as there are leading-order phase space constraints within integrals over next-to-leading order distributions.  This map must be IRC safe, which uniquely constrains the map in the soft or collinear regions of phase space.  However, away from the singular boundaries, this map is ambiguous, and there are many possible choices for such a map.  Perhaps the most natural choices are pairwise jet clustering algorithms, like $k_T$ \cite{Catani:1991hj,Catani:1993hr,Ellis:1993tq} or Cambridge/Aachen \cite{Dokshitzer:1997in,Wobisch:1998wt}, and we will study some dependence on the choice of IRC safe map in \Sec{sec:hbb}.  For now, however, we will leave such a map implicit.

Using the leading-order cumulative distribution of the likelihood, the contribution to the cumulative distribution at next-to-leading order is
\begin{align}\label{eq:cumbacksimp}
\Sigma^{(1)}_B({\cal L})&=\int d\Phi^{(1)}\, p_B^{(1)}(\Phi^{(1)})\left[\Theta\left({\cal L}-\hat{\cal L}^{(0)}(\Phi^{(0)})\right)-\Sigma_B^{(0)}({\cal L})\right]\\
&
\hspace{0.5cm}-{\cal L}\int d\Phi^{(1)} \,
p_B^{(1)}(\Phi^{(1)})\left[\delta\left(
{\cal L}-\hat{\cal L}^{(0)}(\Phi^{(0)})\right) -p_B^{(0)}({\cal L})
\right]\nonumber\\
&\hspace{0.5cm}+{\cal L}^2\int d\Phi^{(1)} \,
p_S^{(1)}(\Phi^{(1)})\left[\delta\left(
{\cal L}-\hat{\cal L}^{(0)}(\Phi^{(0)})\right) -p_S^{(0)}({\cal L})
\right]\nonumber\,,
\end{align}
for background and 
\begin{align}\label{eq:cumsigsimp}
\Sigma^{(1)}_S({\cal L})&=\int d\Phi^{(1)}\, p_S^{(1)}(\Phi^{(1)})\left[\Theta\left({\cal L}-\hat{\cal L}^{(0)}(\Phi^{(0)})\right)-\Sigma_S^{(0)}({\cal L})\right]\\
&
\hspace{0.5cm}+{\cal L}\int d\Phi^{(1)} \,
p_S^{(1)}(\Phi^{(1)})\left[\delta\left(
{\cal L}-\hat{\cal L}^{(0)}(\Phi^{(0)})\right) -p_S^{(0)}({\cal L})
\right]\nonumber\\
&\hspace{0.5cm}-\int d\Phi^{(1)} \,
p_B^{(1)}(\Phi^{(1)})\left[\delta\left(
{\cal L}-\hat{\cal L}^{(0)}(\Phi^{(0)})\right) -p_B^{(0)}({\cal L})
\right]\nonumber\,,
\end{align}
for the signal distribution.  Here, we have introduced the notation for the leading-order probability distribution of the likelihood,
\begin{align}
p^{(0)}({\cal L}) = \frac{d}{d{\cal L}}\,\Sigma^{(0)}({\cal L}) = \int d\Phi^{(0)}\, p^{(0)}(\Phi^{(0)})\,\delta\left({\cal L}-\hat {\cal L}^{(0)}\right)\,.
\end{align}

\subsection{The Inverse Cumulative Distribution of the Likelihood}

Unlike the (cumulative) distribution of the likelihood ratio, there does not exist a simple, direct calculation that can be performed to determine the inverse cumulative distribution, as necessary to evaluate the ROC curve.  However, we can establish its perturbative expansion through its definition as the functional inverse of the cumulative distribution.  Specifically, we require that
\begin{align}
\Sigma^{-1}\left(
\Sigma({\cal L})
\right) = {\cal L}\,,
\end{align}
and we can then match terms in the perturbative expansion order-by-order to satisfy this requirement.

First, we expand the inverse cumulative distribution in powers of $\alpha_s$ as
\begin{align}
\Sigma^{-1}(x) = \Sigma^{-1,(0)}(x)+\frac{\alpha_s}{2\pi}\,\Sigma^{-1,(1)}(x)+{\cal O}(\alpha_s^2)\,.
\end{align}
Next, we take its argument to be the perturbative expansion of the cumulative distribution, where
\begin{align}
\Sigma^{-1}\left(
\Sigma({\cal L})
\right) &= \Sigma^{-1}\left(
\Sigma^{(0)}({\cal L})+\frac{\alpha_s}{2\pi}\, \Sigma^{(1)}({\cal L})+{\cal O}(\alpha_s^2)\right)\\
&=\Sigma^{-1,(0)}\left(\Sigma^{(0)}({\cal L})+\frac{\alpha_s}{2\pi}\, \Sigma^{(1)}({\cal L})\right)+\frac{\alpha_s}{2\pi}\,\Sigma^{-1,(1)}\left(\Sigma^{(0)}({\cal L})\right)+{\cal O}(\alpha_s^2)\nonumber\\
&=\Sigma^{-1,(0)}\left(\Sigma^{(0)}({\cal L})\right)+\left.\frac{\alpha_s}{2\pi}\, \Sigma^{(1)}({\cal L})\,\frac{d}{dx}\Sigma^{-1,(0)}\left(x\right)\right|_{x=\Sigma^{(0)}({\cal L})}+\frac{\alpha_s}{2\pi}\,\Sigma^{-1,(1)}\left(\Sigma^{(0)}({\cal L})\right)\nonumber\\
&\hspace{1cm}+{\cal O}(\alpha_s^2)\,.
\nonumber
\end{align}
By definition, note that
\begin{align}
\Sigma^{-1,(0)}\left(\Sigma^{(0)}({\cal L})\right) = {\cal L}\,,
\end{align}
and so terms at each higher order in $\alpha_s$ must vanish.  In particular, this requires that 
\begin{align}
\Sigma^{-1,(1)}\left(\Sigma^{(0)}({\cal L})\right)=-\left. \Sigma^{(1)}({\cal L})\,\frac{d}{dx}\Sigma^{-1,(0)}\left(x\right)\right|_{x=\Sigma^{(0)}({\cal L})}\,.
\end{align}

By the inverse function theorem, note that the derivative of the inverse cumulative distribution is
\begin{align}
\left.\frac{d}{dx}\Sigma^{-1,(0)}\left(x\right)\right|_{x=\Sigma^{(0)}({\cal L})} = \frac{1}{\frac{d}{d{\cal L}}\Sigma^{(0)}({\cal L})} = \frac{1}{p^{(0)}({\cal L})}\,,
\end{align}
just the inverse of the probability distribution of the likelihood ratio.  Using this result, the inverse cumulative distribution of the likelihood has the perturbative expansion
\begin{align}
\Sigma^{-1}(x) = \Sigma^{-1,(0)}(x)+\frac{\alpha_s}{2\pi}\,\Sigma^{-1,(1)}(x)+{\cal O}(\alpha_s^2)\,,
\end{align}
where the NLO coefficient is
\begin{align}
\Sigma^{-1,(1)}(x) = -\Sigma^{(1)}\left(
\Sigma^{-1,(0)}(x)
\right)\,\frac{1}{p^{(0)}\left(
\Sigma^{-1,(0)}(x)
\right)}\,.
\end{align}
In these expressions, we assume that the leading-order inverse cumulative distribution, $\Sigma^{-1,(0)}(x)$, is straightforward to evaluate.

\subsection{The Perturbative Expansion of the ROC Curve}

We now have all of the ingredients in place to construct the perturbative expansion of the ROC curve for the likelihood.  The ROC curve is, again, 
\begin{align}
\text{ROC}(x) = \Sigma_B\left(
\Sigma_S^{-1}(x)
\right)&=\Sigma_B^{(0)}\left(
\Sigma_S^{-1,(0)}(x)+\frac{\alpha_s}{2\pi}\, \Sigma_S^{-1,(1)}(x)
\right)+\frac{\alpha_s}{2\pi}\, \Sigma_B^{(1)}\left(
\Sigma_S^{-1,(0)}(x)
\right)+{\cal O}(\alpha_s^2)\nonumber\\
&=\Sigma_B^{(0)}\left(
\Sigma_S^{-1,(0)}(x)
\right)+\left.\frac{\alpha_s}{2\pi}\, \Sigma_S^{-1,(1)}(x)\, \frac{d}{d{\cal L}}\Sigma_B^{(0)}\left({\cal L}\right)\right|_{{\cal L} = \Sigma_S^{-1,(0)}(x)}
\nonumber\\
&
\hspace{1cm}+\frac{\alpha_s}{2\pi}\, \Sigma_B^{(1)}\left(
\Sigma_S^{-1,(0)}(x)
\right)+{\cal O}(\alpha_s^2)\,.
\end{align}
Using the results for the inverse cumulative distribution from above, the ROC curve has the expansion
\begin{align}
\text{ROC}(x) &= \Sigma_B^{(0)}\left(
\Sigma_S^{-1,(0)}(x)
\right)+\frac{\alpha_s}{2\pi}\left[ \Sigma_B^{(1)}\left(
\Sigma_S^{-1,(0)}(x)
\right) - \Sigma_S^{(1)}\left(
\Sigma_S^{-1,(0)}(x)
\right)\,\frac{p_B^{(0)}\left(
\Sigma_S^{-1,(0)}(x)
\right)}{p_S^{(0)}\left(
\Sigma_S^{-1,(0)}(x)
\right)}\right]\nonumber\\
&\hspace{1cm}+{\cal O}(\alpha_s^2)\nonumber\\
&=\text{ROC}^{(0)}(x)+\frac{\alpha_s}{2\pi}\,\text{ROC}^{(1)}(x)+{\cal O}(\alpha_s^2)\,.
\end{align}
The NLO contribution to the ROC curve of the likelihood ratio is then
\begin{align}
\text{ROC}^{(1)}(x) = \Sigma_B^{(1)}\left(
\Sigma_S^{-1,(0)}(x)
\right) - \Sigma_S^{(1)}\left(
\Sigma_S^{-1,(0)}(x)
\right)\,\frac{p_B^{(0)}\left(
\Sigma_S^{-1,(0)}(x)
\right)}{p_S^{(0)}\left(
\Sigma_S^{-1,(0)}(x)
\right)}\,.
\end{align}

Using the compact results from the end of \Sec{sec:pertcumlike}, the NLO correction to the ROC curve takes a compact form.  Note that the expression
\begin{align}
\text{ROC}^{(1)}\left(\Sigma_S^{(0)}({\cal L})\right)=\Sigma_B^{(1)}\left(
{\cal L}
\right) - \Sigma_S^{(1)}\left(
{\cal L}
\right)\,\frac{p_B^{(0)}\left(
{\cal L}
\right)}{p_S^{(0)}\left(
{\cal L}
\right)} = \Sigma_B^{(1)}\left(
{\cal L}
\right) -{\cal L}\, \Sigma_S^{(1)}\left(
{\cal L}
\right)\,,
\end{align}
simply from the definition of the likelihood.  With this observation, note from \Eqs{eq:cumbacksimp}{eq:cumsigsimp} that the terms that contain $\delta$-functions explicitly cancel, leaving only the first terms in each, that contain $\Theta$-functions and the leading-order cumulative distributions of the likelihood.  Therefore, the NLO correction to the ROC curve is expressed as
\begin{align}
\text{ROC}^{(1)}\left(\Sigma_S^{(0)}({\cal L})\right) &= \int d\Phi^{(1)}\, p_B^{(1)}(\Phi^{(1)})\left[\Theta\left({\cal L}-\hat{\cal L}^{(0)}(\Phi^{(0)})\right)-\Sigma_B^{(0)}({\cal L})\right]\\
&\hspace{2cm}-{\cal L}\int d\Phi^{(1)}\, p_S^{(1)}(\Phi^{(1)})\left[\Theta\left({\cal L}-\hat{\cal L}^{(0)}(\Phi^{(0)})\right)-\Sigma_S^{(0)}({\cal L})\right]\,.\nonumber
\end{align}

A useful rule-of-thumb for the discrimination power of the likelihood is the area under the ROC curve, or the AUC, where
\begin{align}
\text{AUC} = \int_0^1 dx\, \text{ROC}(x) = \text{AUC}^{(0)}+\frac{\alpha_s}{2\pi}\,\text{AUC}^{(1)}+{\cal O}(\alpha_s^2)\,,
\end{align}
which also admits a perturbative expansion.  The first two contributions are
\begin{align}
\text{AUC}^{(0)} &= \int_0^1 dx\, \Sigma_B^{(0)}\left(
\Sigma_S^{-1,(0)}(x)
\right) = \int_0^\infty d{\cal L}\, p_S^{(0)}({\cal L})\, \Sigma_B^{(0)}({\cal L})\,,\\
\text{AUC}^{(1)} &= \int_0^1dx\, \left(
\Sigma_B^{(1)}\left(
\Sigma_S^{-1,(0)}(x)
\right) - \Sigma_S^{(1)}\left(
\Sigma_S^{-1,(0)}(x)
\right)\,\frac{p_B^{(0)}\left(
\Sigma_S^{-1,(0)}(x)
\right)}{p_S^{(0)}\left(
\Sigma_S^{-1,(0)}(x)
\right)}
\right)\\
&=\int_0^\infty d{\cal L}\,p_S^{(0)}({\cal L})\left(
\Sigma_B^{(1)}({\cal L})-{\cal L}\, \Sigma_S^{(1)}({\cal L})
\right)
\nonumber\,.
\end{align}
We note that the sign of the NLO contribution to the AUC is not well-defined, and so higher-order corrections can either increase or decrease the discrimination power as compared to leading order.  However, note that the AUC will necessarily decrease, corresponding to better discrimination power, if the NLO correction to the background distribution is negative, $\Sigma_B^{(1)}({\cal L})<0$, and the NLO correction to the signal distribution is positive, $\Sigma_S^{(1)}({\cal L})>0$.

\section{Example: Boosted $H\to b\bar b$ from $g\to b\bar b$ Discrimination}\label{sec:hbb}

As a concrete illustration of this framework, we will consider the problem of discrimination of Higgs boson decays from massive gluon splitting to bottom quark pairs, $H\to b\bar b$ from $g\to b\bar b$, in highly-boosted jets.  This problem has a long history in jet physics and a first solution jump started the modern jet substructure program \cite{Butterworth:2008iy}, and has seen numerous studies within the context of machine learning more recently, e.g., \Refs{Lin:2018cin,Datta:2019ndh,Moreno:2019neq,Chakraborty:2019imr,CMS:2020poo,Chung:2020ysf,Tannenwald:2020mhq,Guo:2020vvt,Abbas:2020khd,Jang:2021eph,Khosa:2021cyk}.  It is known that the likelihood for some discrimination problems are not IRC safe, for example, in hadronic decays of boosted top quarks versus massive QCD jets \cite{Soyez:2012hv,Larkoski:2013paa}, or that IRC safety of the likelihood is only realized when all-orders effects are included, for example, in quark versus gluon jet discrimination \cite{Larkoski:2019nwj,Kasieczka:2020nyd}.  However, $H\to b\bar b$ from $g\to b\bar b$ discrimination is IRC safe for the following reasons.  First, the particle content of the final states are identical and so in the likelihood ratio of squared matrix elements divergences associated with addition collinear radiation will exactly cancel.  Second, there is no soft singularity associated with the splitting $g\to b\bar b$, and so its squared matrix element is regular in the soft region of phase space.  Finally, the Higgs has a fixed mass, and so a collinear divergence of the $b\bar b$ in gluon splitting is regulated by a mass cut.

Because the final states of $H\to b\bar b$ and $g\to b\bar b$ are identical and indistinguishable their contributions to the cross section for a given process should, in principle, be summed coherently at the amplitude level.  However, the Higgs has a small decay width compared to its mass $m_H$ and, working in the narrow width approximation, cannot interfere with other intermediate states.  Further, we will work in the highly-boosted limit in which the decay or splitting products are all highly collimated into a small angular region.  In practice, the region of interest would be defined by a jet algorithm, typically anti-$k_T$ \cite{Cacciari:2008gp}, but here, for simplicity and generality, we remain agnostic to the particular jet algorithm employed.  Instead, we work in the approximation that the jet energy $E$ is sufficiently large that all particles produced would be captured by any jet algorithm, to leading order in $m_H^2/E^2$.  Thus, our results will capture the universal behavior of this discrimination problem, and sensitivity to realistic jet energy, jet radius, contamination from initial state radiation, or other effects will be left to future work.  Therefore, this analysis represents an idealized limit of the possible discrimination power, given these simplifications.

Even within this simplified framework, there are other ambiguities that arise at next-to-leading order.  Discrimination of $H\to b\bar b$ decays from $g\to b\bar b$ splittings implicitly assumes that jets have already been tagged as containing at least a bottom and anti-bottom quark.  Modern techniques for bottom hadron tagging have extremely impressive performance (see, e.g., \Refs{CMS:2017wtu,ATLAS:2022qxm}), but there are nevertheless some minimal, reasonable requirements on resolution.  Specifically, the $b$ and $\bar b$ must lie outside some minimal angle from one another to be distinguishable.  At next-to-leading-order, the matrix element for $H\to b\bar b g$ decay has no divergence associated with collinear $b\bar b$, and so any angular cut can be ignored, to leading power in the value of the cut angle.  However, the splitting $g\to b\bar b g$ does have a $b\bar b$ collinear divergence, which correspondingly must be regulated.  Inspired by \Refs{Caletti:2022hnc,Caola:2023wpj}, our procedure here to do this is to simply require that the clustered $b\bar b$ pair has the largest invariant mass, or, equivalently, is clustered by the JADE algorithm \cite{JADE:1986kta,JADE:1988xlj} at the very last step.  This constraint obviously has no effect at leading-order and at next-to-leading order regulates the $b\bar b$ collinear divergence, so will accomplish our needs here.  However, this naive counting of bottomness flavor in each subjet is not IRC safe beyond next-to-leading order \cite{Banfi:2006hf}, and so a generalization would require a well-defined flavor prescription.  There are several other recent examples of IR and IRC safe flavor algorithms \cite{Caletti:2022glq,Czakon:2022wam,Gauld:2022lem}, and we leave a detailed study of their potential consequences for identification of $H\to b\bar b$ decays to future work.

\subsection{Discrimination at Leading Order}

We begin with establishing the baseline discrimination power of these two processes at leading order.  First, working in the highly-boosted limit, we note that leading-order collinear phase space is simply
\begin{align}
d\Phi^{(0)} = \frac{1}{(4\pi)^2}\,ds\, dz\,\delta(s-m_H^2)\,,
\end{align}
where $s$ is the invariant mass of the two final state particles and $z\in[0,1/2]$ is the energy fraction of the softer of the two, assuming otherwise indistinguishability.  We have included the constraint on the mass of the Higgs directly in the phase space.  The squared matrix element for $H\to b\bar b$ decay is flat, and so its normalized probability distribution on phase space is
\begin{align}
p_H^{(0)}(s,z) = 2(4\pi)^2\,,
\end{align}
which integrates to 1 on phase space $d\Phi^{(0)}$ with $z\in[0,1/2]$.

The collinear splitting process $g\to b\bar b$ is governed by the corresponding splitting function \cite{Gribov:1972ri,Gribov:1972rt,Lipatov:1974qm,Dokshitzer:1977sg,Altarelli:1977zs}, represented by the leading-order squared matrix element
\begin{align}
|{\cal M}(g\to b\bar b)|^2 = (4\pi)^2\frac{\alpha_s}{2\pi}\,\frac{1}{s}\,T_R\left(
z^2+(1-z)^2
\right)\,,
\end{align}
where $T_R = 1/2$ is the normalization of the matrices in the fundamental representation of SU(3) color.  We present the complete matrix element because the normalization of the next-to-leading order contribution will depend on the prefactors at leading-order, which we will address in the next section.  The normalized distribution for $g\to b\bar b$ splitting is then
\begin{align}
p_g^{(0)}(s,z) = 3(4\pi)^2\left(
z^2+(1-z)^2
\right)\,,
\end{align}
which integrates to 1 on $d\Phi^{(0)}$.

\begin{figure}
\begin{center}
\includegraphics[width=0.45\textwidth]{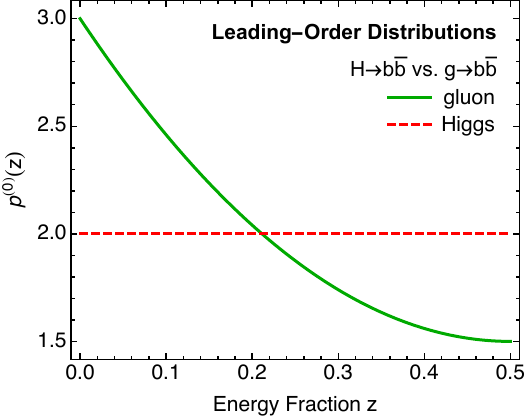}\ \ \ 
\includegraphics[width=0.45\textwidth]{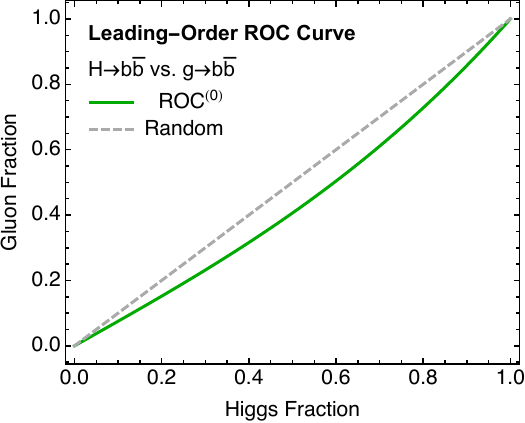}
\caption{\label{fig:lodisc}
Left: Leading-order probability distributions in the energy fraction $z$ for $g\to b\bar b$ splitting (solid green) and $h\to b\bar b$ decay (dashed red).  Right: Leading-order ROC curve of gluon fraction as a function of Higgs fraction (solid green) as compared to random (dashed gray).
}
\end{center}
\end{figure}

The likelihood ratio is then just the ratio between the distributions, where
\begin{align}
\hat {\cal L}^{(0)} = \frac{p_g^{(0)}(s,z)}{p_H^{(0)}(s,z)} = \frac{3}{2} \left(
z^2+(1-z)^2
\right)\,.
\end{align}
Note that this is monotonic on $z\in[0,1/2]$, and so an equally good classifier is the energy fraction $1-z$ itself.  We choose $1-z$ instead of $z$ because both $\hat {\cal L}^{(0)}$ and $1-z$ are decreasing on $z\in[0,1/2]$.  This simplifies the calculation of the ROC curve to this order, as
\begin{align}
\Sigma_H^{(0)}(z) = 2z\,,
\end{align}
and so its inverse is
\begin{align}
\Sigma_H^{-1,(0)}(x) = \frac{x}{2}\,.
\end{align}
The cumulative distribution for the gluon is
\begin{align}
\Sigma_g^{(1)}(z) = \int_0^z dz'\, p_g^{(0)}(s,z') = 2z^3-3z^2+3z\,.
\end{align}
The ROC curve to this order is therefore
\begin{align}
\text{ROC}^{(0)}(x) =1- \Sigma_g^{(0)}\left(
\Sigma_H^{-1,(0)}(1-x)
\right) = \frac{x^3}{4} +\frac{3}{4}\, x\,.
\end{align}
We have plotted the probability distributions and the ROC curve in \Fig{fig:lodisc}.

The leading-order AUC is just as easy to calculate, with
\begin{align}
\text{AUC}^{(0)} = \int_0^1 dx\, \text{ROC}^{(0)}(x) = \int_0^1 dx\, \left(
\frac{x^3}{4} +\frac{3}{4}\, x
\right) = \frac{7}{16}=0.4375\,,
\end{align}
slightly smaller than the AUC of completely random event selection of $0.5$.  Compared to the AUCs for other discrimination problems calculated at lowest order, this is rather large.  For example, the AUC for quark versus gluon jet discrimination at leading-logarithmic accuracy is \cite{Larkoski:2013eya}
\begin{align}
\text{AUC}_\text{$q$ vs.~$g$} = \frac{C_F}{C_F+C_A} = \frac{\frac{4}{3}}{\frac{4}{3}+3}=\frac{4}{13}\approx 0.31\,,
\end{align}
determined by the ratio of the adjoint $C_A$ to the fundamental $C_F$ quadratic Casimirs of SU(3) color.

\subsection{Discrimination at Next-to-Leading Order}

With the analysis at leading-order established, we move to next-to-leading order.  From the general results for the likelihood distributions and ROC curve at NLO from \Sec{sec:rocnlo}, there are two requirements for their practical calculation.  First, we need an IRC safe map from NLO to LO phase spaces, so we can evaluate the leading-order cumulative distribution on the distribution of particles at next-to-leading order.  Second, while the likelihood ratio is IRC safe, there will be divergences at intermediate steps of the calculation that must be appropriately isolated and subtracted, to render the NLO contribution finite.  We will describe our approach to both of these issues and present numerical results for the ROC curve for $H\to b\bar b$ from $g\to b\bar b$ discrimination at next-to-leading order in this section.

\subsubsection{Map from NLO to LO Phase Space}

For the NLO-to-LO phase space map, we need NLO phase space in the first place.  In the highly-boosted limit with fixed total invariant mass, differential three-body phase space can be expressed as \cite{Gehrmann-DeRidder:1997fom,Ritzmann:2014mka}
\begin{align}
d\Phi^{(1)} &\equiv \frac{4}{(4\pi)^{5}} \frac{ds_{b\bar b}\, ds_{bg}\, ds_{\bar b g}\, dz_b\, dz_{\bar b}\, dz_g}{\sqrt{4z_bz_{\bar b} s_{bg}s_{\bar b g}-(z_g s_{b\bar b}-z_b s_{\bar b g}-z_{\bar b}s_{bg})^2}}\\
&
\hspace{3cm}\times \delta\left(
m_H^2-s_{b\bar b}-s_{bg}-s_{\bar b g}
\right)\delta(1-z_b-z_{\bar b}-z_g)\nonumber\,.
\end{align}
Here, $s_{ij}$ is the invariant mass of particles $i$ and $j$ and $z_i$ is the energy fraction of particle $i$.  We will present the explicit form of the squared NLO matrix elements soon, but even without them, we already know their infrared divergent structure and therefore what the NLO-to-LO map must regulate.  Because we enforce a lower bound on the invariant mass of the $b\bar b$ pair, they have no collinear divergence, and so the only divergences are when the gluon becomes collinear to either the $b$ or $\bar b$, or when the gluon becomes soft/low energy.  In the collinear limits, this NLO-to-LO map must cluster collinear particles first for IRC safety.

Thus, the simplest such NLO-to-LO map is just that from a sequential jet clustering algorithm, and here we will focus on the $k_T$ family of algorithms.  Specifically, we will study the Cambridge/Aachen (C/A) \cite{Dokshitzer:1997in,Wobisch:1998wt} algorithm.  With the additional JADE algorithm requirement on the $b\bar b$ pair, the C/A mapping can be expressed in these phase space coordinates as
\begin{align}
\delta^{(1)}_\text{C/A}(z)&\equiv 
\Theta\left(s_{b\bar b}-\min[s_{\bar b g},s_{\bar b g}]\right)\left[\Theta\left(\frac{s_{bg}}{z_b}-\frac{s_{\bar b g}}{z_{\bar b}}\right)\,\delta\left(
z-\min[z_b,1-z_b]
\right)\right.\\
&\hspace{6cm}\left.+\Theta\left(\frac{s_{\bar bg}}{z_{\bar b}}-\frac{s_{b g}}{z_b}\right)\,\delta\left(
z-\min[z_{\bar b},1-z_{\bar b}]
\right)\right]\nonumber\,.
\end{align} 
Here, $z$ is the energy fraction of the softer of the two subjets after reclustering, which is also the one non-trivial LO phase space coordinate.  Again, because any possible collinear divergence of the $b\bar b$ pair has been regulated by JADE, we do not need consider clustering them together to return finite results at NLO.  The clustering constraints can be expressed in the general form where
\begin{align}
\delta_\text{alg}^{(1)}(z) &= \Theta\left(s_{b\bar b}-\min[s_{\bar b g},s_{\bar b g}]\right)\Theta\left(g_\text{alg}(z_b,z_g)s_{bg}-g_\text{alg}(z_{\bar b},z_g)s_{\bar b g}\right)\delta\left(
z-\min[z_b,1-z_b]
\right)\nonumber\\
&
\hspace{2cm}+(b\leftrightarrow \bar b)\,,
\end{align}
where $g_\text{alg}$ is a function of energy fractions that determines the specific form of the IRC safe clustering algorithm.

\subsubsection{Subtraction Method for Numerical NLO Evaluation}\label{sec:subsec}

From the expression for the cumulative distribution of the likelihood ratio in \Eq{eq:likecumdist}, the basic object that appears at next-to-leading order takes the form of either
\begin{align}\label{eq:baseob}
\int d\Phi^{(1)}\, p^{(1)}(\Phi^{(1)})\left[\Theta\left({\cal L}-\hat{\cal L}^{(0)}(\Phi^{(0)})\right)-\Sigma^{(0)}({\cal L})\right]\,,
\end{align}
or its derivative with respect to ${\cal L}$.  The leading-order cumulative distributions can be directly evaluated using the results of the previous section.  We have
\begin{align}
\Sigma_H^{(0)}({\cal L})=\int d\Phi^{(0)}\,p_H^{(0)}(\Phi^{(0)}) \,\Theta\left(
{\cal L}-\hat {\cal L}^{(0)}
\right) &=\sqrt{\frac{4}{3}{\cal L}-1}\,,\\
\Sigma_g^{(0)}({\cal L})=\int d\Phi^{(0)}\,p_g^{(0)}(\Phi^{(0)}) \,\Theta\left(
{\cal L}-\hat {\cal L}^{(0)}
\right) &=\left(
\frac{1}{2}+\frac{{\cal L}}{3}
\right)\sqrt{\frac{4}{3}{\cal L}-1}\,,
\end{align}
for ${\cal L}\in[3/4,3/2]$.  As it is by itself IRC safe, we will consider \Eq{eq:baseob} as the fundamental object we will calculate at NLO, and then from it, construct the full NLO cumulative distributions according to the prescriptions established earlier.

The phase space constraints of the NLO-to-LO map are complicated enough that there is no hope to evaluate this analytically, so we must do at least some of the phase space integrals numerically.  IRC safety, however, is not enough to ensure that intermediate steps of a calculation do not have divergences that ultimately cancel at the end.  Thus, as written with the expected divergences in the NLO distribution, \Eq{eq:baseob} can not be naively numerically evaluated.  We can, then, add and subtract to each integral a term that exactly cancels the divergences, which we denote as $p^{(1,\text{sing})}(\Phi^{(1)})$, in an analogous way to familiar dipole subtraction methods, like ERT \cite{Ellis:1980wv}, FKS \cite{Frixione:1995ms}, or Catani-Seymour \cite{Catani:1996vz} methods. We now have
\begin{align}
&\int d\Phi^{(1)}\, p^{(1)}(\Phi^{(1)})\left[\Theta\left({\cal L}-\hat{\cal L}^{(0)}(\Phi^{(0)})\right)-\Sigma^{(0)}({\cal L})\right]\\\
&
\hspace{1cm} = \int d\Phi^{(1)}\, \left(p^{(1)}(\Phi^{(1)})-p^{(1,\text{sing})}(\Phi^{(1)})\right) \left[\Theta\left(
{\cal L}-\hat {\cal L}^{(0)}
\right) -\Sigma^{(0)}({\cal L})\right]\nonumber\\
&
\hspace{2cm}+\int d\Phi^{(1)}\,p^{(1,\text{sing})}(\Phi^{(1)}) \left[\Theta\left(
{\cal L}-\hat {\cal L}^{(0)}
\right)-\Sigma^{(0)}({\cal L})\right] \nonumber\,.
\end{align}
Now, the first integrand is explicitly finite over all of phase space and can be evaluated with standard Monte Carlo methods.  Much of the integral on the final line, exclusively with the singular subtraction matrix element, can be evaluated exactly analytically in dimensional regularization and the divergences explicitly removed.  For brevity and presentation clarity here, we present the details of this calculation in \App{app:singints}.

For the signal and background processes considered here, as well as the requirement on $b$-tagging, the leading-power singular distribution takes the form
\begin{align}\label{eq:singdistnlo}
p^{(1,\text{sing})}(\Phi^{(1)}) &= (4\pi)^2\,p^{(0)}(\Phi^{(0)})\left[
\frac{C_F}{s_{\bar b g}}\left(\frac{1+\frac{z_{\bar b}^2}{(z_{\bar b}+z_g)^2}}{\frac{z_g}{z_{\bar b}+z_g}}-\frac{2}{\frac{z_g}{z_{\bar b}}}\right)\right.\\
&
\hspace{4cm}\left.+\frac{C_F}{s_{b g}}\left(\frac{1+\frac{z_{b}^2}{(z_{b}+z_g)^2}}{\frac{z_g}{z_{b}+z_g}}-\frac{2}{\frac{z_g}{z_{b}}}\right)+p^{(1,\text{soft})}(\Phi^{(1)})
\right]\nonumber\\
&=(4\pi)^2\,p^{(0)}(\Phi^{(0)})\left[
\frac{C_F}{s_{\bar b g}}\,\frac{z_g}{z_{\bar b}+z_g}+\frac{C_F}{s_{b g}}\,\frac{z_g}{z_b+z_g}+p^{(1,\text{soft})}(\Phi^{(1)})
\right]\nonumber\,.
\end{align}
Here, we have explicitly subtracted the soft-collinear region from both the $bg$ and $\bar b g$ collinear regions, leaving it in the soft function contribution.  There is no $b\bar b$ collinear region because of the flavor tagging constraints. The soft contribution depends on the flow of color in the initial state, where, for the $H\to b\bar b$ and $g\to b\bar b$ processes, we have, respectively,
\begin{align}
p_H^{(1,\text{soft})}(\Phi^{(1)}) &= 2C_F\,\frac{s_{b\bar b}}{s_{bg}s_{\bar b g}}\,,\\
p_g^{(1,\text{soft})}(\Phi^{(1)}) &= 2C_F\,\frac{s_{b\bar b}}{s_{bg}s_{\bar b g}}+C_A\,
\frac{z_bs_{\bar b g}+z_{\bar b}s_{b g}-z_g s_{b\bar b}}{z_gs_{bg}s_{\bar b g}}\,.
\end{align}

\subsubsection{NLO Matrix Elements}

The final piece to establish to be able to numerically evaluate the likelihood distributions at NLO are the squared matrix elements at NLO themselves.  For Higgs decay, the real, three-body decay matrix element can be straightforwardly calculated to be
\begin{align}
\frac{|{\cal M}(H\to b\bar b g)|^2}{|{\cal M}(H\to b\bar b)|^2} = (4\pi)^2\,\frac{\alpha_s}{2\pi}\,C_F\,\frac{1}{m_H^2} \frac{s_{b\bar b}^2+m_H^4}{s_{bg}s_{\bar b g}}\,.
 \end{align}
This relationship to the leading-order matrix element then means that the properly-normalized contribution to the phase space distribution at NLO is 
\begin{align}
p^{(1)}_H(\Phi^{(1)}) = 2(4\pi)^4\,C_F\,\frac{1}{m_H^2} \frac{s_{b\bar b}^2+m_H^4}{s_{bg}s_{\bar b g}}\,.
\end{align}
$C_F=4/3$ is the fundamental quadratic Casimir of SU(3) color.  Note, of course, that this matrix element is Lorentz-invariant and applies in any frame, including the highly-boosted collinear frame of our analysis.

For high-energy gluon splitting, the matrix element at leading-power in the collinear limit takes the general form
\begin{align}
|{\cal M}(g\to b\bar b g)|^2 = (4\pi)^4\left(\frac{\alpha_s}{2\pi}\right)^2\frac{T_R}{m_H^4}\left(
C_F\, P^\text{ab}_{g\to b\bar b g}(\Phi^{(1)})+C_A\, P^\text{nab}_{g\to b\bar b g}(\Phi^{(1)})
\right)\,,
\end{align}
where $P^\text{ab}_{g\to b\bar b g}(\Phi^{(1)})$ and $P^\text{nab}_{g\to b\bar b g}(\Phi^{(1)})$ are the Abelian and non-Abelian splitting functions, as defined by their color prefactor.  Here, $C_A=3$ is the adjoint quadratic Casimir of SU(3) color.  With the normalization of the leading-order distribution, the properly-normalized NLO contribution is then
\begin{align}
p_g^{(1)}(\Phi^{(1)}) = 3(4\pi)^4\frac{1}{m_H^2}\left(
C_F\, P^\text{ab}_{g\to b\bar b g}(\Phi^{(1)})+C_A\, P^\text{nab}_{g\to b\bar b g}(\Phi^{(1)})
\right)\,.
\end{align}

The individual collinear splitting functions are \cite{Campbell:1997hg,Catani:1999ss}
\begin{align}
P^\text{ab}_{g\to b\bar b g}(\Phi^{(1)}) &= -2-s_{b\bar b}\left(
\frac{1}{s_{bg}}+\frac{1}{s_{\bar b g}}\right)+\frac{2m_H^4}{s_{bg}s_{\bar b g}}\left(
1+z_g^2-z_g-2z_bz_{\bar b}
\right)\\
&\hspace{7cm}-\frac{m_H^2}{s_{bg}}\left(
1-2z_b
\right)-\frac{m_H^2}{s_{\bar bg}}\left(
1-2z_{\bar b}
\right)\nonumber\,,\\
P^\text{nab}_{g\to b\bar b g}(\Phi^{(1)})&=-\frac{1}{2s_{b\bar b}^2}\left(
2\frac{z_bs_{\bar b g}-z_{\bar b}s_{bg}}{z_b+z_{\bar b}}+\frac{z_b-z_{\bar b}}{z_b+z_{\bar b}}\,s_{b\bar b}
\right)^2\\
&
\hspace{-1cm}+\frac{m_H^4}{2s_{b\bar b}}\left(\frac{z_{\bar b}}{s_{\bar b g}}\,
\frac{(1-z_g)^3-z_g^3-2z_{\bar b}(1-z_{\bar b}-2z_gz_b)}{z_g(1-z_g)}+\frac{z_{b}}{s_{b g}}\,
\frac{(1-z_g)^3-z_g^3-2z_{b}(1-z_{b}-2z_gz_{\bar b})}{z_g(1-z_g)}\right)\nonumber\\
&\hspace{-1cm}+\frac{(1-z_b)m_H^2}{2s_{\bar b g}}\,\frac{z_g(1-z_g)+1-2z_b(1-z_b)}{z_g(1-z_g)}+\frac{(1-z_{\bar b})m_H^2}{2s_{b g}}\,\frac{z_g(1-z_g)+1-2z_{\bar b}(1-z_{\bar b})}{z_g(1-z_g)}\nonumber\\
&\hspace{-1cm}+\frac{m_H^2}{s_{b\bar b}}\,\frac{1+z_g^3+z_g(z_b-z_{\bar b})^2-2z_bz_{\bar b}(1+z_g)}{z_g(1-z_g)}-\frac{1}{2}-\frac{m_H^4}{s_{bg}s_{\bar b g}}\left(
1+z_g^2-z_g-2z_bz_{\bar b}
\right)\,.\nonumber
\end{align}
Unlike the Higgs decay matrix element, these splitting functions only apply in the highly-boosted regime and have explicit dependence on non-Lorentz invariant quantities of the particle energy fractions.  This is of course because the gluon is massless and any non-zero mass ascribed to the gluon only arises because of its high-energy fragmentation.

One issue with the non-Abelian splitting function that must be dealt with is the fact that, in addition to its physical infrared divergences, it also has an unphysical ``collinear'' divergence when the angle of the gluon to either the $b$ or $\bar b$ diverges, $\theta_{bg}^2,\theta_{\bar bg}^2\to\infty$, with fixed $s_{bg},s_{\bar b g}$.  This divergences exists because the color octet gluon must recoil against the rest of the event that itself carries a net octet color, and so the gluon can be preferentially emitted collinear to the rest of the event.  Within our high-boost, collinear approximation, this is of course not accurately described and anyway there will always be some finite jet radius that separates the jet from the rest of the event.  Our solution here to regulate this divergence is therefore to impose a maximal angle cut on the emitted gluon, where we demand that the squared angle of the gluon from either the $b$ or $\bar b$ is no larger than 8 times the squared angle between the $b\bar b$ itself:
\begin{align}
8\,\theta_{b\bar b}^2>\max[\theta_{bg}^2,\theta_{\bar bg}^2]\,.
\end{align}
For reference, for jets in experiments with radius $R\sim 1$, this roughly corresponds to identification of Higgs bosons with an energy about $5$ to 6 times its mass, $E \sim 700$ GeV.  At any rate, the results are relatively weakly dependent on the particular value of this cut; varying it by a factor of two changes the values of the integrals at NLO by about 10\% (i.e., the infinite angle divergence is logarithmic).  Additionally, because the Abelian splitting function proportional to $C_F$ is integrable as angles diverge, this constraint has a negligible effect on that component of the matrix element.

In the truly infinite boost limit, but finite, non-zero jet radius, the angle between the $b$ and $\bar b$ goes to 0.  As the Higgs is a color singlet, radiation from the $b\bar b$ dipole would also be restricted to a region of 0 angular size, assuming there is no contamination emitted from outside of the jet.  By contrast, because the gluon is a color-octet, it would still emit radiation at all angles throughout the jet.  Thus, strictly speaking, in the infinite-boost limit, there would be sufficient information for {\it perfect} discrimination between Higgs decays versus gluon fragmentation to bottom quarks.  This distinction of radiation patterns in the highly-boosted limit was recently exploited in construction of the jet color ring observable \cite{Buckley:2020kdp}.  Therefore, for realistic and useful predictions for discrimination, we must assume that the energy of the jet is finite.  As mentioned above, the dependence on discrimination power with the boost is logarithmic, and so exhibits mild variation over experimentally-accessible jet energies.

\subsection{Numerical Results}

\begin{figure}[t!]
\begin{center}
\includegraphics[width=0.45\textwidth]{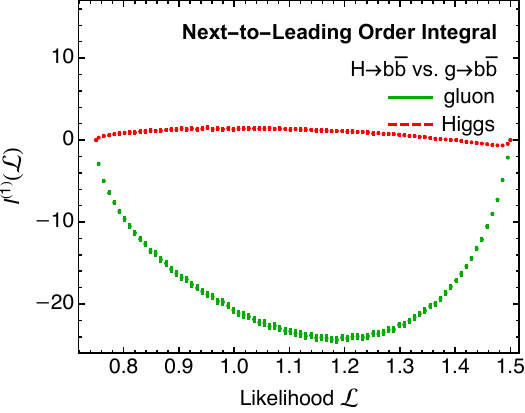}
\caption{\label{fig:baseobNLO}
Plots of the basic integral $I^{(1)}({\cal L})$ appearing at next-to-leading order for $H\to b\bar b$ from $g\to b\bar b$ discrimination, defined in \Eq{eq:baseint_nlo}, as a function of the leading-order likelihood ratio ${\cal L}$.  The C/A algorithm is used to map NLO- to LO-phase space.  Error bars represent uncertainty on the numerical integration.
}
\end{center}
\end{figure}

With these accumulated results, we can evaluate all necessary integrals to establish the cumulative distribution of the likelihood ratio on Higgs and gluon events.  We perform numerical integration with the implementation of VEGAS \cite{Lepage:1977sw,Lepage:1980dq} in the CUBA libraries \cite{Hahn:2004fe}.  The first thing that we show is the functional form of the basic object appearing at next-to-leading order; namely
\begin{align}\label{eq:baseint_nlo}
I^{(1)}({\cal L})=\int d\Phi^{(1)}\, p^{(1)}(\Phi^{(1)})\left[\Theta\left({\cal L}-\hat{\cal L}^{(0)}(\Phi^{(0)})\right)-\Sigma^{(0)}({\cal L})\right]\,,
\end{align}
on both Higgs and gluon events.  These are displayed in \Fig{fig:baseobNLO} over the range ${\cal L}\in[3/4,3/2]$, the range of the leading-order likelihood ratio.  There are a few properties to note; first, and most importantly, both integrals vanish at the endpoints of ${\cal L}=3/4,3/2$, which ensures that the probability distributions of the likelihood ratio through NLO remains unit normalized. Next, note that the gluon integral is negative over the domain of the likelihood, while the Higgs integral is positive over (nearly all) of the domain.  Thus, in the probability distributions of the likelihood ratios proportional to the derivatives of these integrals, effects at NLO tend to push the gluon distribution to larger ${\cal L}$ (where its derivative is positive) and the Higgs distribution to smaller ${\cal L}$ (where its derivative is positive).  Thus, NLO effects tend to improve the discrimination power between these classes of events.

From this basic object, we can then calculate the cumulative distributions of the likelihood ratio through next-to-leading order.  To do this requires evaluation of the strong coupling $\alpha_s(\mu^2)$ at a scale $\mu^2$.  Because we consider on-shell Higgs decays, the natural scale is $\mu^2 = m_H^2$, just the Higgs mass itself.  The established value of the strong coupling evaluated at the $Z$ pole is $\alpha_s(m_Z^2) = 0.1179$ \cite{ParticleDataGroup:2022pth}, and with one-loop running, the value of the strong coupling at the Higgs mass is therefore $\alpha_s(m_H^2) = 0.113$.  A widespread practice for estimation of theoretical uncertainties is to vary this scale by a factor of 2, and then consider the sensitivity to this variation as quantifying theoretical error by truncation of the perturbative expansion in $\alpha_s$.  At this scale, this scale sensitivity is roughly a $10\%$ effect of the NLO correction.  In the plots we present, we will show central values with $\alpha_s = 0.113$, and bands about this value where we have varied the scale of the coupling by a factor of 2.

\begin{figure}[t!]
\begin{center}
\includegraphics[width=0.45\textwidth]{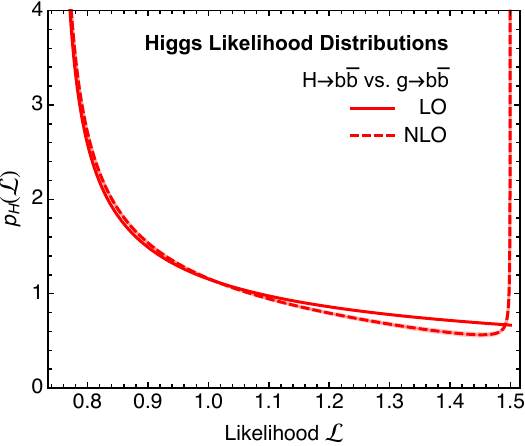}\ \ \ 
\includegraphics[width=0.45\textwidth]{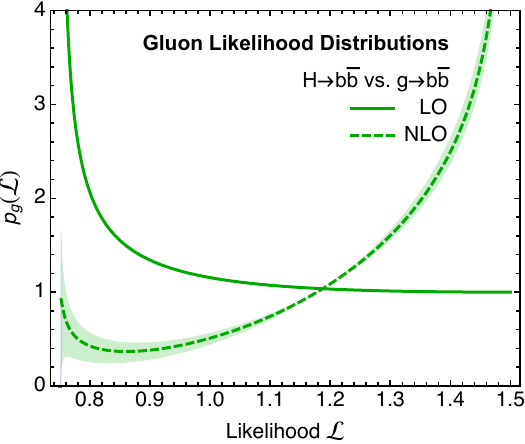}
\caption{\label{fig:cumdistNLO}
Plots of the probability distribution of the likelihood ratio ${\cal L}$ at leading- (solid) and next-to-leading (dashed) order for Higgs events (left) and gluon events (right), both clustered with the C/A algorithm.  At next-to-leading order, we use the central value $\alpha_s(m_H^2) = 0.113$, and the shaded bands represents sensitivity to variation of the scale of the coupling by a factor of 2.  
}
\end{center}
\end{figure}

We plot the probability distributions of the likelihood ratio through next-to-leading order in \Fig{fig:cumdistNLO}.  On each plot, we show both the leading- and next-to-leading order distributions for Higgs events at left and gluons at right.  As relevant for discrimination and what appears in the ROC curve from the discussion in \Sec{sec:pertcumlike}, in evaluation of the distributions at NLO, we only include the contributions from \Eqs{eq:cumbacksimp}{eq:cumsigsimp} which contain the $\Theta$-functions (that is, contributions of the form of the basic integral $I^{(1)}({\cal L})$).  Note that the modification from NLO effects to the likelihood distribution on Higgs events is rather small, and comparable to the size of effects of variation of the scale of the coupling.  However, NLO effects to the gluon distribution are extremely large, because only first at NLO is there radiation that probes the flow of color in the jets.  The radiation pattern of gluons is significantly distinct from color-singlet Higgs bosons especially at wide angles, and this feature significantly improves discrimination power.  We do expect, however, that contributions from next-to-next-to-leading order and higher are relatively small modifications of the NLO distribution because there are no qualitatively new physical effects that arise.

\begin{figure}[t!]
\begin{center}
\includegraphics[width=0.45\textwidth]{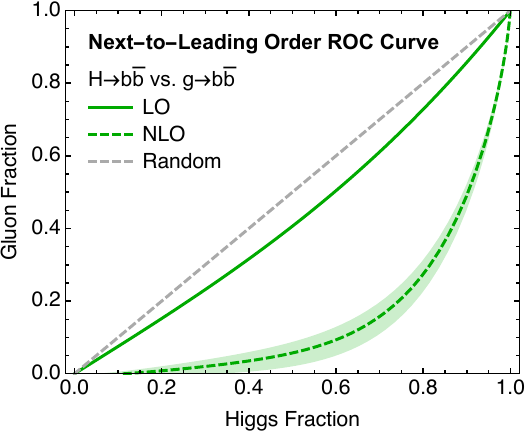}
\caption{\label{fig:nlorocs}
Plots of the ROC curves for Higgs decays to bottom quarks  vs.~gluon splitting to bottom quarks at leading (solid) and next-to-leading (dashed) orders.  At next-to-leading order, we use the central value $\alpha_s(m_H^2) = 0.113$, and the shaded bands represents sensitivity to variation of the scale of the coupling by a factor of 2.  A random classifier is represented by the line in dashed gray.
}
\end{center}
\end{figure}

From these likelihood distributions, we can then evaluate the corresponding ROC curves through next-to-leading order.  This is plotted in \Fig{fig:nlorocs}.  This unambiguously demonstrates that discrimination is dramatically improved at NLO, and is starting to be comparable to results from machine learning a classifier on simulated data (see, e.g., \Refnew{Khosa:2021cyk}).  To quantify the discrimination power in a single number, we can evaluate the AUC through next-to-leading order, where we find
\begin{align}
\text{AUC} = \frac{7}{16}-15.7\,\frac{\alpha_s}{2\pi}+{\cal O}(\alpha_s^2)\,.
\end{align}
The NLO coefficient, $-15.7$, is the value of the numerical integral of the NLO contribution to the ROC curve.  With the physical value of the coupling at the Higgs mass, the AUC through NLO is approximately
\begin{align}
\text{AUC}= 0.156 \pm 0.025\,,
\end{align}
where the uncertainty is representative of the sensitivity to the scale of the coupling.

\section{Conclusions}\label{sec:conc}

For binary classification problems in jet physics that admit a perturbative expansion in the strong coupling, $\alpha_s$, we have formulated discrimination with the likelihood ratio through next-to-leading order.  Through this analysis, we have explicitly identified the properties that are necessary for an infrared and collinear safe classifier, and applied the general results to the concrete example of discrimination of Higgs decays from gluon splitting to bottom quarks in the highly-boosted limit.  Next-to-leading order effects are vital for any qualitative understanding of the classifier because that is the first order at which sensitivity to the flow of color in the jets appears.  In the true infinite boost limit and ignoring contamination, the classifier is perfect because only gluons will emit radiation at finite angles from the $b\bar b$ pair.  At high but accessible boosts that can be measured at the Large Hadron Collider, these radiation effects are substantial, and decrease the area under the ROC curve by about a factor of 3 from the leading-order prediction.  Only starting at next-to-leading order are these predictions qualitatively close to corresponding results from machine learning studies on simulated data.

This property also demonstrates that predictions at next-to-next-to-leading order (NNLO) are necessary for an accurate uncertainty estimate.  This is similar to giant $K$-factors that arise in studies of vector boson production \cite{Butterworth:2008iy,Bauer:2009km,Denner:2009gj,Rubin:2010xp}.  In that case, naive scale variation fails to account for next-to-leading order effects because, starting at next-to-leading order, there are qualitatively new processes that contribute, corresponding to distinct initial states, whose presence cannot be estimated by simple scale variation alone.  Here, the ``giant $K$-factor'' arises because, starting at next-to-leading order, there are qualitatively new diagrams corresponding to the presence of radiation off of the hard, Born-level particles.  Within the framework of this highly-boosted analysis, one could continue by including contributions from four-particle final states, as represented by, for example, $1\to 4$ splitting functions.  While these have been recently calculated \cite{DelDuca:2019ggv,DelDuca:2020vst}, it would be simpler and likely a good approximation to just use strongly-ordered splitting functions, like employed here in \Eq{eq:singdistnlo} for practical numerical integration.  The importance of these higher-order contributions would then establish the robustness of discrimination at NLO.

While the decay of the Higgs to bottoms quarks has been observed \cite{ATLAS:2020bhl,ATLAS:2020fcp,CMS:2022dwd}, the decay of Higgs bosons to gluons has not, even though its branching fraction is about $10\%$.  $H\to gg$ shares many features with $H\to b\bar b$ decays, but the lack of bottom quarks that can be identified and tagged means that backgrounds for $H\to gg$ are overwhelmingly large.  Even though gluons from Higgs decay must be in a color-singlet combination, their larger individual color charge renders their corresponding jets ``fuzzier'' than that of quarks, and standard observables sensitive to the flow of color are much less sensitive \cite{Gallicchio:2010sw,Curtin:2012rm,Buckley:2020kdp}.  At least for discrimination of $H\to gg$ decays from $g\to gg$ splitting, the IRC safe likelihood analysis here could provide necessary insight into both strict upper bounds on performance as well as new, useful observables for this problem.  However, $g\to gg$ splitting has a soft divergence, unlike $g\to b\bar b$, and thus, even at leading order, there will be logarithmic sensitivity to the jet energy and radius that is responsible for regulating the soft, wide-angle divergence.

While many discrimination problems do not admit a likelihood ratio that is IRC safe, such as hadronic top quark decay \cite{Soyez:2012hv}, this analysis may seem of limited utility.  However, in the top quark decay example, many of the leading-order final states initiated by QCD jets are dramatically different from the top.  For example, while the decay $t\to b q\bar q'$ can look experimentally identical to, say, $g\to b\bar b g$ when the $\bar b$ is mistagged, theoretically, these final states are distinguishable. This may suggest that a way forward to establish ultimate theoretical performance by identification of background subprocesses that produce a final state as close as possible to that of signal.  Assuming perfect bottom hadron tagging, there is only one process initiated by light(er) QCD partons that mimics top decay; namely, $b\to b q\bar q$, the high-energy fragmentation of a bottom quark to three quarks.  Both the top and its daughter $W$ boson decay on-shell in the narrow-width approximation, and correspondingly imposing the top mass constraint on the background final state $b q\bar q$ and the $W$ mass on the sub-state $q\bar q$ completely eliminates the presence of infrared divergences and renders the likelihood ratio IRC safe.  A theoretical analysis of top tagging with this approach might then quantitatively establish a hard upper bound on performance, and explain the saturation of discrimination from machine learning \cite{Kasieczka:2019dbj}.

\acknowledgments

I thank Gregor Kasieczka for discussions and providing the motivation for this analysis and Simone Marzani for comments.  This work was supported in part by the UC Southern California Hub, with funding from the UC National Laboratories division of the University of California Office of the President.

\appendix

\section{Integral Over Singular Subtraction Terms}\label{app:singints}

The integral over the singular subtraction terms necessary to evaluate the cumulative distribution of the likelihood ratio at next-to-leading order as discussed in \Sec{sec:subsec} takes the form
\begin{align}
I^{(1,\text{sing})}\equiv\int d\Phi^{(1)}\,p^{(1,\text{sing})}(\Phi^{(1)}) \left[\Theta\left(
{\cal L}-\hat {\cal L}^{(0)}
\right)-\Sigma^{(0)}({\cal L})\right]\,.
\end{align}
In this appendix, we will perform this integral over phase space within dimensional regularization where $d=4-2\epsilon$ and so infrared divergences can be explicitly isolated and subtracted.  Three-body collinear phase space in $d=4-2\epsilon$ dimensions is \cite{Gehrmann-DeRidder:1997fom,Ritzmann:2014mka}
\begin{align}
d\Phi^{(1)} &= \frac{4}{(4\pi)^{5-2\epsilon}\Gamma(1-2\epsilon)}\, \frac{ds_{b\bar b}\, ds_{bg}\, ds_{\bar b g}\, dz_b\, dz_{\bar b}\, dz_g}{\left(4z_bz_{\bar b} s_{bg}s_{\bar b g}-(z_g s_{b\bar b}-z_b s_{\bar b g}-z_{\bar b} s_{bg})^2\right)^{\frac{1}{2}+\epsilon}}\\
&
\hspace{5cm}\times \delta\left(
m_H^2-s_{b\bar b}-s_{bg}-s_{\bar bg}
\right)\,\delta(1-z_b-z_{\bar b}-z_g)\nonumber
\end{align}
The singular distribution takes the general form
\begin{align}
p^{(1,\text{sing})}(\Phi^{(1)}) 
&=(4\pi)^2\,p^{(0)}(\Phi^{(0)})\left[
\frac{C_F}{s_{\bar b g}}\,\frac{z_g}{z_{\bar b}+z_g}+\frac{C_F}{s_{b g}}\,\frac{z_g}{z_b+z_g}+p^{(1,\text{soft})}(\Phi^{(1)})
\right]\,,
\end{align}
where the soft contributions depend on the flow of color from the initiating particle, where
\begin{align}
p_H^{(1,\text{soft})}(\Phi^{(1)}) &= 2C_F\,\frac{s_{b\bar b}}{s_{bg}s_{\bar b g}}\,,\\
p_g^{(1,\text{soft})}(\Phi^{(1)}) &= 2C_F\,\frac{s_{b\bar b}}{s_{bg}s_{\bar b g}}+C_A\,
\frac{z_bs_{\bar b g}+z_{\bar b}s_{b g}-z_g s_{b\bar b}}{z_gs_{bg}s_{\bar b g}}\,.
\end{align}
The general form of the IRC safe map from NLO to LO phase space is
\begin{align}
\delta^{(1)}_\text{alg}(z) &= \Theta\left(s_{b\bar b}-\min[s_{b g},s_{\bar b g}]\right)\Theta\left(g_\text{alg}(z_b,z_g)s_{bg}-g_\text{alg}(z_{\bar b},z_g)\, s_{\bar b g}\right)\delta\left(
z-\min[z_b,1-z_b]
\right)\nonumber\\
&
\hspace{-1cm}
+\Theta\left(s_{b\bar b}-\min[s_{bg},s_{\bar bg}]\right)\, \Theta\left(g_\text{alg}(z_{\bar b},z_g)\,s_{\bar bg}-g_\text{alg}(z_b,z_g)\, s_{b g}\right)\delta\left(
z-\min[z_{\bar b},1-z_{\bar b}]
\right)
\end{align}

To actually perform the map from NLO to LO in the integration, we also need to factorize NLO phase space into it LO and emission component.  For a fixed invariant mass, collinear two-body phase space in dimensional regularization is \cite{Giele:1991vf}
\begin{align}
d\Phi^{(0)} = \frac{1}{(4\pi)^{2-\epsilon}}\frac{(m_H^2)^{-\epsilon}}{\Gamma(1-\epsilon)}\,z^{-\epsilon}(1-z)^{-\epsilon}\, dz\,
\end{align}
where $z\in[0,1]$ is the energy fraction of one  of the two particles at LO.  Phase space can then be factorized as
\begin{align}
d\Phi^{(1)} &= d\Phi^{(0)}\, \frac{4(m_H^2)^{\epsilon}\,\Gamma(1-\epsilon)}{(4\pi)^{3-\epsilon}\Gamma(1-2\epsilon)}\, \frac{ds_{b\bar b}\, ds_{bg}\, ds_{\bar b g}\, dz_b\, dz_{\bar b}\, dz_g\, z^{\epsilon}\,(1-z)^\epsilon}{\left(4z_bz_{\bar b} s_{bg}s_{\bar b g}-(z_g s_{b\bar b}-z_b s_{\bar b g}-z_{\bar b} s_{bg})^2\right)^{\frac{1}{2}+\epsilon}}\\
&
\hspace{5cm}\times \delta\left(
m_H^2-s_{b\bar b}-s_{bg}-s_{\bar bg}
\right)\,\delta(1-z_b-z_{\bar b}-z_g)\, \delta^{(1)}_\text{alg}(z)\,.
\nonumber\\
&\equiv d\Phi^{(0)}\,\frac{d\Phi^{(1)}}{d\Phi^{(0)}}\, \delta^{(1)}_\text{alg}(z)\nonumber\,.
\end{align}

Note that everything is symmetric in $b\leftrightarrow \bar b$ so we can write
\begin{align}
I^{(1,\text{sing})}({\cal L}) &=2\int d\Phi^{(0)} \left[\Theta\left(
{\cal L}-\hat {\cal L}^{(0)}(\Phi^{(0)})
\right)-\Sigma^{(0)}({\cal L})\right] \int \frac{d\Phi^{(1)}}{d\Phi^{(0)}}\, p^{(1,\text{sing})}(\Phi^{(1)})\\
&\hspace{0.5cm}\times\Theta\left(s_{b\bar b}-\min[s_{b g},s_{\bar b g}]\right)\Theta\left(g_\text{alg}(z_b,z_g)s_{bg}-g_\text{alg}(z_{\bar b},z_g)\, s_{\bar b g}\right)\delta\left(
z-\min[z_b,1-z_b]
\right)\nonumber\,.
\end{align}
With this phase space factorization, all divergences will be isolated to the integral over the pure emission phase space.  To isolate and subtract divergences, we therefore evaluate $d\Phi^{(1)}/d\Phi^{(0)}$ in $4-2\epsilon$ dimensions, but once divergences are subtracted, the integral over LO phase space can be evaluated in $d=4$ dimensions.  Explicitly, this is then
\begin{align}
I^{(1,\text{sing})}({\cal L}) &=\frac{2}{(4\pi)^2}\int dz \left[\Theta\left(
{\cal L}-\hat {\cal L}^{(0)}(z)
\right)-\Sigma^{(0)}({\cal L})\right] \int \frac{d\Phi^{(1)}}{d\Phi^{(0)}}\, p^{(1,\text{sing})}(\Phi^{(1)})\\
&\hspace{0.5cm}\times\Theta\left(s_{b\bar b}-\min[s_{b g},s_{\bar b g}]\right)\Theta\left(g_\text{alg}(z_b,z_g)s_{bg}-g_\text{alg}(z_{\bar b},z_g)\, s_{\bar b g}\right)\delta\left(
z-\min[z_b,1-z_b]
\right)\nonumber\,.
\end{align}

We will consider each term in the expression for the singular distribution separately and then sum together their contributions at the end.  We will calculate complete, IR divergence-subtracted contributions and will demonstrate that all contributions are explicitly IRC safe.

\subsection{Finite Term}

With these clustering constraints, the integrand of one term is finite on all of the available phase space:
\begin{align}
I_1^{(1,\text{sing})}({\cal L}) &\equiv 2(4\pi)^2\,C_F\int_0^{1/2} dz\, p^{(0)}(z) \left[\Theta\left(
{\cal L}-\hat {\cal L}^{(0)}(z)
\right)-\Sigma^{(0)}({\cal L})\right] \int d\Phi^{(1)}\, \frac{1}{s_{b g}}\,\frac{z_g}{z_b+z_g}\nonumber\\
&\hspace{0.5cm}\times\Theta\left(s_{b\bar b}-\min[s_{b g},s_{\bar b g}]\right)\Theta\left(g_\text{alg}(z_b,z_g)s_{bg}-g_\text{alg}(z_{\bar b},z_g)\, s_{\bar b g}\right)\delta\left(
z-\min[z_b,1-z_b]
\right)\,.
\end{align}
This can therefore be evaluated with usual Monte Carlo methods.

\subsection{Collinear Divergent Term}

One term exclusively has a single collinear divergence, where
\begin{align}
I_2^{(1,\text{sing})}({\cal L}) &\equiv 2\,C_F\int dz\, p^{(0)}(z) \left[\Theta\left(
{\cal L}-\hat {\cal L}^{(0)}(z)
\right)-\Sigma^{(0)}({\cal L})\right] \int \frac{d\Phi^{(1)}}{d\Phi^{(0)}}\, \frac{1}{s_{\bar b g}}\,\frac{z_g}{z_{\bar b}+z_g}\\
&\hspace{0.5cm}\times\Theta\left(s_{b\bar b}-\min[s_{b g},s_{\bar b g}]\right)\Theta\left(g_\text{alg}(z_b,z_g)s_{bg}-g_\text{alg}(z_{\bar b},z_g)\, s_{\bar b g}\right)\delta\left(
z-\min[z_b,1-z_b]
\right)\,.\nonumber
\end{align}
To isolate and subtract the divergences, we will re-write phase space in angle-energy fraction coordinates for which
\begin{align}
\frac{d\Phi^{(1)}}{d\Phi^{(0)}} &= \frac{4^{1-\epsilon}(m_H^2)^\epsilon(E^2)^{-2\epsilon}E^2}{(4\pi)^{3-\epsilon}}\frac{\Gamma(1-\epsilon)}{\Gamma(1-2\epsilon)}\,d\theta_{\bar b g}^2\,d\theta^2_{b\bar b}\, d\phi\,dz_{\bar b}\,dz_g\,z_b^{1-\epsilon}\,(1-z_b)^\epsilon\,z_{\bar b}^{1-2\epsilon}\,z_g^{1-2\epsilon}(\theta_{\bar b g}^2)^{-\epsilon}(\theta_{b\bar b}^2)^{-\epsilon}\sin^{-2\epsilon}\phi\nonumber\\
&\hspace{2cm}
\times\delta\left(
\frac{m_H^2}{E^2}-z_bz_{\bar b}\theta_{b\bar b}^2-z_{\bar b}z_g\theta_{\bar b g}^2-z_bz_g(\theta_{b\bar b}^2+\theta_{\bar b g}^2-2\theta_{b\bar b}\theta_{\bar b g}\cos\phi)
\right)\\
&\hspace{2cm}\times \delta(1-z_b-z_{\bar b}-z_g)\,.\nonumber
\end{align}
Here, $E$ is the energy of the jet, $\phi$ is azimuthal angle of the gluon about the line that passes through the $b\bar b$ pair, and we have used the law of cosines in the invariant mass $\delta$-function, where
\begin{align}
\theta_{bg}^2=\theta_{b\bar b}^2+\theta_{\bar b g}^2-2\theta_{b\bar b}\theta_{\bar b g}\cos\phi\,.
\end{align}
In these coordinates, the distribution and clustering constraints are
\begin{align}
&\frac{1}{s_{\bar b g}}\,\frac{z_g}{z_{\bar b}+z_g}\,\Theta\left(s_{b\bar b}-\min[s_{b g},s_{\bar b g}]\right)\Theta\left(g_\text{alg}(z_b,z_g)s_{bg}-g_\text{alg}(z_{\bar b},z_g)\, s_{\bar b g}\right)\\
&=\frac{1}{E^2z_{\bar b}z_g\theta_{\bar b g}^2}\,\frac{z_g}{z_{\bar b}+z_g}\,\Theta\left(z_bz_{\bar b}\theta_{b\bar b}^2-\min[z_bz_g\theta_{bg}^2,z_{\bar b}z_g\theta_{\bar b g}^2]\right)\Theta\left(g_\text{alg}(z_b,z_g)\, z_b\theta_{bg}^2-g_\text{alg}(z_{\bar b},z_g)\,z_{\bar b}\theta_{\bar b g}^2\right)\nonumber\,.
\end{align}
We have left the angle $\theta_{bg}^2$ implicit here for compactness.  Also, note that we have explicitly fixed the LO phase space coordinate to be the energy fraction $z_b$.

Combining these results, the integral over this phase space is
\begin{align}
&\int \frac{d\Phi^{(1)}}{d\Phi^{(0)}}\, p^{(1,\text{sing})}(\Phi^{(1)})\\
&\supset \frac{4^{1-\epsilon}(m_H^2)^\epsilon(E^2)^{-2\epsilon}}{(4\pi)^{3-\epsilon}}\frac{\Gamma(1-\epsilon)}{\Gamma(1-2\epsilon)}\int  \frac{d\theta_{\bar b g}^2}{(\theta_{\bar b g}^2)^{1+\epsilon}}\,d\theta^2_{b\bar b}\, d\phi\,dz_{\bar b}\,dz_g\,\frac{z_b^{1-\epsilon}\,z_{\bar b}^{-2\epsilon}\,z_g^{1-2\epsilon}}{(1-z_b)^{1-\epsilon}}\,(\theta_{b\bar b}^2)^{-\epsilon}\sin^{-2\epsilon}\phi\nonumber\\
&\hspace{1cm}
\times\delta\left(
\frac{m_H^2}{E^2}-z_bz_{\bar b}\theta_{b\bar b}^2-z_{\bar b}z_g\theta_{\bar b g}^2-z_bz_g(\theta_{b\bar b}^2+\theta_{\bar b g}^2-2\theta_{b\bar b}\theta_{\bar b g}\cos\phi)
\right)\delta(1-z_b-z_{\bar b}-z_g)\nonumber\\
&\hspace{1cm}\times \Theta\left(z_bz_{\bar b}\theta_{b\bar b}^2-\min[z_bz_g\theta_{bg}^2,z_{\bar b}z_g\theta_{\bar b g}^2]\right)\Theta\left(g_\text{alg}(z_b,z_g)\, z_b\theta_{bg}^2-g_\text{alg}(z_{\bar b},z_g)\,z_{\bar b}\theta_{\bar b g}^2\right)\,.
\nonumber
\end{align}
The collinear divergence is now explicit.  We can isolate the divergence using the $+$-function expansion, where
\begin{align}
\frac{1}{(\theta_{\bar b g}^2)^{1+\epsilon}} = -\frac{1}{\epsilon}\left(
\frac{m_H^2}{E^2}
\right)^{-\epsilon}\delta(\theta_{\bar b g}^2)+\left(
\frac{1}{\theta_{\bar b g}^2}
\right)_+-\epsilon\left(
\frac{\log \theta_{\bar b g}^2}{\theta_{\bar b g}^2}
\right)_++\cdots\,.
\end{align}
The $+$-functions are defined to integrate to 0 on $[0,m_H^2/E^2]$, where
\begin{align}
0 = \int_0^{m_H^2/E^2}dx\, f_+(x)\,,
\end{align}
and when integrated over a function $g(x)$ that is analytic about $x=0$ is defined as
\begin{align}
\int_0^{m_H^2/E^2} dx\, f_+(x)\, g(x) = \int_0^{m_H^2/E^2}dx\, f(x)\left[
g(x)-g(0)
\right]\,.
\end{align}

With this expansion, we can then perform the integration, only keeping those terms necessary to capture through order-$\epsilon^0$.  We have
\begin{align}
&\int \frac{d\Phi^{(1)}}{d\Phi^{(0)}}\, p^{(1,\text{sing})}(\Phi^{(1)})\\
&\supset \frac{4^{1-\epsilon}(m_H^2)^\epsilon(E^2)^{-2\epsilon}}{(4\pi)^{3-\epsilon}}\frac{\Gamma(1-\epsilon)}{\Gamma(1-2\epsilon)}\int  d\theta_{\bar b g}^2\left(
-\frac{1}{\epsilon}\left(
\frac{m_H^2}{E^2}
\right)^{-\epsilon}\delta(\theta_{\bar b g}^2)+\left(
\frac{1}{\theta_{\bar b g}^2}
\right)_+
\right)\,d\theta^2_{b\bar b}\, d\phi\,dz_{\bar b}\,dz_g\nonumber\\
&\hspace{1cm}
\times\,\frac{z_b^{1-\epsilon}\,z_{\bar b}^{-2\epsilon}\,z_g^{1-2\epsilon}}{(1-z_b)^{1-\epsilon}}\,(\theta_{b\bar b}^2)^{-\epsilon}\sin^{-2\epsilon}\phi\nonumber\\
&\hspace{1cm}\times\delta\left(
\frac{m_H^2}{E^2}-z_bz_{\bar b}\theta_{b\bar b}^2-z_{\bar b}z_g\theta_{\bar b g}^2-z_bz_g(\theta_{b\bar b}^2+\theta_{\bar b g}^2-2\theta_{b\bar b}\theta_{\bar b g}\cos\phi)
\right)\delta(1-z_b-z_{\bar b}-z_g)\nonumber\\
&\hspace{1cm}\times \Theta\left(z_bz_{\bar b}\theta_{b\bar b}^2-\min[z_bz_g\theta_{bg}^2,z_{\bar b}z_g\theta_{\bar b g}^2]\right)\Theta\left(g_\text{alg}(z_b,z_g)\, z_b\theta_{bg}^2-g_\text{alg}(z_{\bar b},z_g)\,z_{\bar b}\theta_{\bar b g}^2\right)\,.
\nonumber
\end{align}
The divergent integral can then be explicitly evaluated to find
\begin{align}
&
-\frac{4^{1-\epsilon}(E^2)^{-\epsilon}}{(4\pi)^{3-\epsilon}}\frac{\Gamma(1-\epsilon)}{\Gamma(1-2\epsilon)}\frac{1}{\epsilon}\int  d\theta_{\bar b g}^2\,\delta(\theta_{\bar b g}^2)\,d\theta^2_{b\bar b}\, d\phi\,dz_{\bar b}\,dz_g\,\frac{z_b^{1-\epsilon}\,z_{\bar b}^{-2\epsilon}\,z_g^{1-2\epsilon}}{(1-z_b)^{1-\epsilon}}\,(\theta_{b\bar b}^2)^{-\epsilon}\sin^{-2\epsilon}\phi\nonumber\\
&\hspace{1cm}\times\delta\left(
\frac{m_H^2}{E^2}-z_bz_{\bar b}\theta_{b\bar b}^2-z_{\bar b}z_g\theta_{\bar b g}^2-z_bz_g(\theta_{b\bar b}^2+\theta_{\bar b g}^2-2\theta_{b\bar b}\theta_{\bar b g}\cos\phi)
\right)\delta(1-z_b-z_{\bar b}-z_g)\nonumber\\
&\hspace{1cm}\times \Theta\left(z_bz_{\bar b}\theta_{b\bar b}^2-\min[z_bz_g\theta_{bg}^2,z_{\bar b}z_g\theta_{\bar b g}^2]\right)\Theta\left(g_\text{alg}(z_b,z_g)\, z_b\theta_{bg}^2-g_\text{alg}(z_{\bar b},z_g)\,z_{\bar b}\theta_{\bar b g}^2\right)\nonumber\\
&=\frac{(m_H^2)^{-\epsilon}}{(4\pi)^{2-\epsilon}}\left(
-\frac{1}{2\epsilon}+\log(1-z_b)-2
\right)\,.
\end{align}
Note that the residual scale dependence establishes that the natural scale at which to evaluate $\alpha_s(\mu^2)$ is at the Higgs mass, $\mu^2 = m_H^2$.

The integral with the $+$-function can be massaged into the form
\begin{align}
&\frac{4}{(4\pi)^3}\int  d\theta_{\bar b g}^2\left(
\frac{1}{\theta_{\bar b g}^2}
\right)_+\,d\theta^2_{b\bar b}\, d\phi\,dz_{\bar b}\,dz_g\,\frac{z_b\,z_g}{(1-z_b)}\nonumber\\
&\hspace{1cm}\times\delta\left(
\frac{m_H^2}{E^2}-z_bz_{\bar b}\theta_{b\bar b}^2-z_{\bar b}z_g\theta_{\bar b g}^2-z_bz_g(\theta_{b\bar b}^2+\theta_{\bar b g}^2-2\theta_{b\bar b}\theta_{\bar b g}\cos\phi)
\right)\delta(1-z_b-z_{\bar b}-z_g)\nonumber\\
&\hspace{1cm}\times \Theta\left(z_bz_{\bar b}\theta_{b\bar b}^2-\min[z_bz_g\theta_{bg}^2,z_{\bar b}z_g\theta_{\bar b g}^2]\right)\Theta\left(g_\text{alg}(z_b,z_g)\, z_b\theta_{bg}^2-g_\text{alg}(z_{\bar b},z_g)\,z_{\bar b}\theta_{\bar b g}^2\right)\\
&=\frac{4}{(4\pi)^3}\int  
\frac{d\theta_{\bar b g}^2}{\theta_{\bar b g}^2}\,d\theta^2_{b\bar b}\, d\phi\,dz_{\bar b}\,dz_g\,\frac{z_b\,z_g}{(1-z_b)}\,\delta(1-z_b-z_{\bar b}-z_g)\,\Theta\left(
\frac{m_H^2}{E^2}-\theta_{\bar b g}^2
\right)\nonumber\\
&\hspace{1cm}\times\left[\delta\left(
\frac{m_H^2}{E^2}-z_bz_{\bar b}\theta_{b\bar b}^2-z_{\bar b}z_g\theta_{\bar b g}^2-z_bz_g(\theta_{b\bar b}^2+\theta_{\bar b g}^2-2\theta_{b\bar b}\theta_{\bar b g}\cos\phi)
\right)\right.\nonumber\\
&\hspace{1cm}\left.\times \Theta\left(z_bz_{\bar b}\theta_{b\bar b}^2-\min[z_bz_g\theta_{bg}^2,z_{\bar b}z_g\theta_{\bar b g}^2]\right)\Theta\left(g_\text{alg}(z_b,z_g)\, z_b\theta_{bg}^2-g_\text{alg}(z_{\bar b},z_g)\,z_{\bar b}\theta_{\bar b g}^2\right)-\delta\left(
\frac{m_H^2}{E^2}-z_b(1-z_z)\theta_{b\bar b}^2
\right)\right]
\nonumber\\
&+\frac{4}{(4\pi)^3}\int  
\frac{d\theta_{\bar b g}^2}{\theta_{\bar b g}^2}\,d\theta^2_{b\bar b}\, d\phi\,dz_{\bar b}\,dz_g\,\frac{z_b\,z_g}{(1-z_b)}\,\delta(1-z_b-z_{\bar b}-z_g)\,\Theta\left(
\theta_{\bar b g}^2-\frac{m_H^2}{E^2}
\right)\nonumber\\
&\hspace{1cm}\times\delta\left(
\frac{m_H^2}{E^2}-z_bz_{\bar b}\theta_{b\bar b}^2-z_{\bar b}z_g\theta_{\bar b g}^2-z_bz_g(\theta_{b\bar b}^2+\theta_{\bar b g}^2-2\theta_{b\bar b}\theta_{\bar b g}\cos\phi)
\right)\nonumber\\
&\hspace{1cm}\times \Theta\left(z_bz_{\bar b}\theta_{b\bar b}^2-\min[z_bz_g\theta_{bg}^2,z_{\bar b}z_g\theta_{\bar b g}^2]\right)\Theta\left(g_\text{alg}(z_b,z_g)\, z_b\theta_{bg}^2-g_\text{alg}(z_{\bar b},z_g)\,z_{\bar b}\theta_{\bar b g}^2\right)
\nonumber\,.
\end{align}
Both integrals are now explicitly finite and can be integrated with standard Monte Carlo methods.  Further, angles can be rescaled by $m_H^2/E^2$ to eliminate any explicit dependence on either the Higgs mass or the jet energy.

Putting these results together and explicitly subtracting divergences, we find that
\begin{align}
&I_2^{(1,\text{sing})}({\cal L}) =2 C_F\int_0^{1/2} dz\, p^{(0)}(z) \left[\Theta\left(
{\cal L}-\hat {\cal L}^{(0)}(z)
\right)-\Sigma^{(0)}({\cal L})\right]\\
&\times\Biggl[
\frac{\log(1-z)+\log z}{(4\pi)^2}\nonumber\\
&\hspace{1cm}+\frac{4}{(4\pi)^3}\int  
\frac{d\theta_{\bar b g}^2}{\theta_{\bar b g}^2}\,d\theta^2_{b\bar b}\, d\phi\,dz_b\,dz_{\bar b}\,dz_g\,\frac{z_b\,z_g}{(1-z_b)}\,\delta(1-z_b-z_{\bar b}-z_g)\,\,\delta\left(
z-\min[z_b,1-z_b]
\right)\nonumber\\
&\hspace{2cm}\times\left[\delta\left(
\frac{m_H^2}{E^2}-z_bz_{\bar b}\theta_{b\bar b}^2-z_{\bar b}z_g\theta_{\bar b g}^2-z_bz_g(\theta_{b\bar b}^2+\theta_{\bar b g}^2-2\theta_{b\bar b}\theta_{\bar b g}\cos\phi)
\right)\right.\nonumber\\
&\hspace{3cm}\times \Theta\left(z_bz_{\bar b}\theta_{b\bar b}^2-\min[z_bz_g\theta_{bg}^2,z_{\bar b}z_g\theta_{\bar b g}^2]\right)\Theta\left(g_\text{alg}(z_b,z_g)\, z_b\theta_{bg}^2-g_\text{alg}(z_{\bar b},z_g)\,z_{\bar b}\theta_{\bar b g}^2\right)\nonumber\\
&\hspace{4cm}\left.-\,\delta\left(
\frac{m_H^2}{E^2}-z_b(1-z_z)\theta_{b\bar b}^2
\right)\Theta\left(
\frac{m_H^2}{E^2}-\theta_{\bar b g}^2
\right)\right]\nonumber\,.
\end{align}

\subsection{Abelian Soft Term}

The Abelian soft term, the term that would exist in electromagnetism, is
\begin{align}
I_3^{(1,\text{sing})}({\cal L}) &\equiv 4\,C_F\int_0^{1/2} dz\, p^{(0)}(z) \left[\Theta\left(
{\cal L}-\hat {\cal L}^{(0)}(z)
\right)-\Sigma^{(0)}({\cal L})\right] \int \frac{d\Phi^{(1)}}{d\Phi^{(0)}}\, \frac{s_{b\bar b}}{s_{bg}s_{\bar b g}}\\
&\hspace{0.5cm}\times\Theta\left(s_{b\bar b}-\min[s_{b g},s_{\bar b g}]\right)\Theta\left(g_\text{alg}(z_b,z_g)s_{bg}-g_\text{alg}(z_{\bar b},z_g)\, s_{\bar b g}\right)\delta\left(
z-\min[z_b,1-z_b]
\right)\nonumber\,.
\end{align}
This has both soft and collinear divergences and so will need to be regulated both in the angle $\theta_{\bar b g}^2$ and the energy fraction $z_g$.  To do this, we again express phase space in angle-energy fraction coordinates.  We than have
\begin{align}
&\int \frac{d\Phi^{(1)}}{d\Phi^{(0)}}\, p^{(1,\text{sing})}(\Phi^{(1)})\\
&\supset  \frac{4^{1-\epsilon}(m_H^2)^\epsilon(E^2)^{-2\epsilon}}{(4\pi)^{3-\epsilon}}\frac{\Gamma(1-\epsilon)}{\Gamma(1-2\epsilon)}\int  \frac{d\theta_{\bar b g}^2}{(\theta_{\bar b g}^2)^{1+\epsilon}}\,d\theta^2_{b\bar b}\, d\phi\,dz_{\bar b}\,\frac{dz_g}{z_g^{1+2\epsilon}}\,z_b^{1-\epsilon}\,(1-z_b)^{\epsilon}\,z_{\bar b}^{1-2\epsilon}\,\frac{(\theta_{b\bar b}^2)^{1-\epsilon}\sin^{-2\epsilon}\phi}{\theta_{b\bar b}^2+\theta_{\bar b g}^2-2\theta_{b\bar b}\theta_{\bar b g}\cos\phi}\nonumber\\
&\hspace{1cm}
\times\delta\left(
\frac{m_H^2}{E^2}-z_bz_{\bar b}\theta_{b\bar b}^2-z_{\bar b}z_g\theta_{\bar b g}^2-z_bz_g(\theta_{b\bar b}^2+\theta_{\bar b g}^2-2\theta_{b\bar b}\theta_{\bar b g}\cos\phi)
\right)\delta(1-z_b-z_{\bar b}-z_g)\nonumber\\
&\hspace{1cm}\times \Theta\left(z_bz_{\bar b}\theta_{b\bar b}^2-\min[z_bz_g\theta_{bg}^2,z_{\bar b}z_g\theta_{\bar b g}^2]\right)\Theta\left(g_\text{alg}(z_b,z_g)\, z_b\theta_{bg}^2-g_\text{alg}(z_{\bar b},z_g)\,z_{\bar b}\theta_{\bar b g}^2\right)\delta\left(
z-\min[z_b,1-z_b]
\right)\,.
\nonumber
\end{align}
We have already discussed the $+$-function expansion for the angle $\theta_{\bar b g}^2$; we now introduce the $+$-function expansion for the energy fraction.  We have the expansion
\begin{align}
\frac{1}{z_g^{1+2\epsilon}} = -\frac{(1-z_b)^{-2\epsilon}}{2\epsilon}\,\delta(z_g)+\left(
\frac{1}{z_g}
\right)_+-2\epsilon\left(
\frac{\log z_g}{z_g}
\right)_++\cdots\,,
\end{align}
and the $+$-functions are defined to integrate to 0 on $z_g\in[0,1-z_b]$.  That is, for an analytic function $g(z_g)$, we define
\begin{align}
\int_0^{1-z_b}dz_g\, f_+(z_g)\,g(z_g) = \int_0^{1-z_b}dz_g\, f(z_g)\left[
g(z_g)-g(0)
\right]\,.
\end{align}

Using the $+$-functions, the integral becomes
\begin{align}
&\int \frac{d\Phi^{(1)}}{d\Phi^{(0)}}\, p^{(1,\text{sing})}(\Phi^{(1)})\\
&\supset   \frac{4^{1-\epsilon}(m_H^2)^\epsilon(E^2)^{-2\epsilon}}{(4\pi)^{3-\epsilon}}\frac{\Gamma(1-\epsilon)}{\Gamma(1-2\epsilon)}\int  d\theta_{\bar b g}^2\left(
-\frac{1}{\epsilon}\left(
\frac{m_H^2}{E^2}
\right)^{-\epsilon}\delta(\theta_{\bar b g}^2)
\right)\,d\theta^2_{b\bar b}\, d\phi\,dz_{\bar b}\,\frac{dz_g}{z_g^{1+2\epsilon}}\,z_b^{1-\epsilon}\,(1-z_b)^{\epsilon}\,z_{\bar b}^{1-2\epsilon}\,\nonumber\\
&\hspace{1cm}
\times\frac{(\theta_{b\bar b}^2)^{1-\epsilon}\sin^{-2\epsilon}\phi}{\theta_{b\bar b}^2+\theta_{\bar b g}^2-2\theta_{b\bar b}\theta_{\bar b g}\cos\phi}\,\delta\left(
\frac{m_H^2}{E^2}-z_bz_{\bar b}\theta_{b\bar b}^2-z_{\bar b}z_g\theta_{\bar b g}^2-z_bz_g(\theta_{b\bar b}^2+\theta_{\bar b g}^2-2\theta_{b\bar b}\theta_{\bar b g}\cos\phi)
\right)\nonumber\\
&\hspace{1cm}\times\delta(1-z_b-z_{\bar b}-z_g) \Theta\left(z_bz_{\bar b}\theta_{b\bar b}^2-\min[z_bz_g\theta_{bg}^2,z_{\bar b}z_g\theta_{\bar b g}^2]\right)\Theta\left(g_\text{alg}(z_b,z_g)\, z_b\theta_{bg}^2-g_\text{alg}(z_{\bar b},z_g)\,z_{\bar b}\theta_{\bar b g}^2\right)\nonumber\\
&\hspace{1cm}\times\delta\left(
z-\min[z_b,1-z_b]
\right)\nonumber\\
&+\frac{4^{1-\epsilon}(m_H^2)^\epsilon(E^2)^{-2\epsilon}}{(4\pi)^{3-\epsilon}}\frac{\Gamma(1-\epsilon)}{\Gamma(1-2\epsilon)}\int  \frac{d\theta_{\bar b g}^2}{(\theta_{\bar b g}^2)^{1+\epsilon}}\,d\theta^2_{b\bar b}\, d\phi\,dz_{\bar b}\,dz_g\left(
-\frac{(1-z_b)^{-2\epsilon}}{2\epsilon}\,\delta(z_g)
\right)z_b^{1-\epsilon}\,(1-z_b)^{\epsilon}\,z_{\bar b}^{1-2\epsilon}\,\nonumber\\
&\hspace{1cm}
\times\frac{(\theta_{b\bar b}^2)^{1-\epsilon}\sin^{-2\epsilon}\phi}{\theta_{b\bar b}^2+\theta_{\bar b g}^2-2\theta_{b\bar b}\theta_{\bar b g}\cos\phi}\,\delta\left(
\frac{m_H^2}{E^2}-z_bz_{\bar b}\theta_{b\bar b}^2-z_{\bar b}z_g\theta_{\bar b g}^2-z_bz_g(\theta_{b\bar b}^2+\theta_{\bar b g}^2-2\theta_{b\bar b}\theta_{\bar b g}\cos\phi)
\right)\nonumber\\
&\hspace{1cm}\times\delta(1-z_b-z_{\bar b}-z_g)\, \Theta\left(z_bz_{\bar b}\theta_{b\bar b}^2-\min[z_bz_g\theta_{bg}^2,z_{\bar b}z_g\theta_{\bar b g}^2]\right)\Theta\left(g_\text{alg}(z_b,z_g)\, z_b\theta_{bg}^2-g_\text{alg}(z_{\bar b},z_g)\,z_{\bar b}\theta_{\bar b g}^2\right)\nonumber\\
&\hspace{1cm}\times\delta\left(
z-\min[z_b,1-z_b]
\right)\nonumber\\
&-\frac{4^{1-\epsilon}(m_H^2)^\epsilon(E^2)^{-2\epsilon}}{(4\pi)^{3-\epsilon}}\frac{\Gamma(1-\epsilon)}{\Gamma(1-2\epsilon)}\int  d\theta_{\bar b g}^2\left(
-\frac{1}{\epsilon}\left(
\frac{m_H^2}{E^2}
\right)^{-\epsilon}\delta(\theta_{\bar b g}^2)
\right)\,d\theta^2_{b\bar b}\, d\phi\,dz_{\bar b}\,dz_g\left(
-\frac{(1-z_b)^{-2\epsilon}}{2\epsilon}\,\delta(z_g)
\right)\nonumber\\
&\hspace{1cm}
\times z_b^{1-\epsilon}\,(1-z_b)^{\epsilon}\,z_{\bar b}^{1-2\epsilon}\,\frac{(\theta_{b\bar b}^2)^{1-\epsilon}\sin^{-2\epsilon}\phi}{\theta_{b\bar b}^2+\theta_{\bar b g}^2-2\theta_{b\bar b}\theta_{\bar b g}\cos\phi}\,\delta(1-z_b-z_{\bar b}-z_g)\, \Theta\left(z_bz_{\bar b}\theta_{b\bar b}^2-\min[z_bz_g\theta_{bg}^2,z_{\bar b}z_g\theta_{\bar b g}^2]\right)\nonumber\\
&\hspace{1cm}\times\delta\left(
\frac{m_H^2}{E^2}-z_bz_{\bar b}\theta_{b\bar b}^2-z_{\bar b}z_g\theta_{\bar b g}^2-z_bz_g(\theta_{b\bar b}^2+\theta_{\bar b g}^2-2\theta_{b\bar b}\theta_{\bar b g}\cos\phi)
\right)\Theta\left(g_\text{alg}(z_b,z_g)\, z_b\theta_{bg}^2-g_\text{alg}(z_{\bar b},z_g)\,z_{\bar b}\theta_{\bar b g}^2\right)\nonumber\\
&\hspace{1cm}\times\delta\left(
z-\min[z_b,1-z_b]
\right)\nonumber\\
&+\frac{4}{(4\pi)^3}\int  d\theta_{\bar b g}^2\left(
\frac{1}{\theta_{\bar b g}^2}
\right)_+\,d\theta^2_{b\bar b}\, d\phi\,dz_{\bar b}\,dz_g\left(
\frac{1}{z_g}
\right)_+\,z_b\,z_{\bar b}\,\frac{\theta_{b\bar b}^2}{\theta_{b\bar b}^2+\theta_{\bar b g}^2-2\theta_{b\bar b}\theta_{\bar b g}\cos\phi}\nonumber\\
&\hspace{1cm}
\times \delta(1-z_b-z_{\bar b}-z_g)\, \Theta\left(z_bz_{\bar b}\theta_{b\bar b}^2-\min[z_bz_g\theta_{bg}^2,z_{\bar b}z_g\theta_{\bar b g}^2]\right)\delta\left(
z-\min[z_b,1-z_b]
\right)\nonumber\\
&\hspace{1cm}\times\delta\left(
\frac{m_H^2}{E^2}-z_bz_{\bar b}\theta_{b\bar b}^2-z_{\bar b}z_g\theta_{\bar b g}^2-z_bz_g(\theta_{b\bar b}^2+\theta_{\bar b g}^2-2\theta_{b\bar b}\theta_{\bar b g}\cos\phi)
\right)\Theta\left(g_\text{alg}(z_b,z_g)\, z_b\theta_{bg}^2-g_\text{alg}(z_{\bar b},z_g)\,z_{\bar b}\theta_{\bar b g}^2\right)\nonumber\,.
\end{align}
The third contribution subtracts the overlap of the double pole in $\epsilon$ from the first two contributions.  For $k_T$-type algorithms that we consider here, clustering with 0 energy particles reduces to a restriction on relative angles exclusively.  That is, 
\begin{align}
\Theta\left(g_\text{alg}(z_b,z_g=0)\, z_b\theta_{bg}^2-g_\text{alg}(z_{\bar b},z_g=0)\,z_{\bar b}\theta_{\bar b g}^2\right)&=\Theta\left(\theta_{bg}^2-\theta_{\bar b g}^2\right)\\
&=\Theta\left(\theta_{b\bar b}-2\theta_{\bar b g}\cos\phi\right)\nonumber\,.
\end{align}
Performing the integrals with the $\delta$-functions, we have
\begin{align}
&\int \frac{d\Phi^{(1)}}{d\Phi^{(0)}}\, p^{(1,\text{sing})}(\Phi^{(1)})\\
&\supset   \frac{(m_H^2)^{-\epsilon}}{(4\pi)^{2-\epsilon}}\left(\frac{1}{\epsilon}+4-\frac{\pi^2}{3}-2\log(1-z_b)\right) \delta\left(
z-\min[z_b,1-z_b]
\right)\nonumber\\
&-\frac{4^{1-\epsilon}(E^2)^{-\epsilon}}{(4\pi)^{3-\epsilon}}\frac{\Gamma(1-\epsilon)}{\Gamma(1-2\epsilon)}\frac{1}{2\epsilon}\int  \frac{d\theta_{\bar b g}^2}{(\theta_{\bar b g}^2)^{1+\epsilon}}\, d\phi\,(1-z_b)^{-2\epsilon}\,\frac{m_H^2\,\sin^{-2\epsilon}\phi}{m_H^2+z_b(1-z_b)E^2\theta_{\bar b g}^2-2\sqrt{z_b(1-z_b)}\,E\,m_H\theta_{\bar b g}\cos\phi}\nonumber\\
&\hspace{1cm}
\times \Theta\left(\frac{m_H}{\sqrt{z_b(1-z_b)}\,E}-2\theta_{\bar b g}\cos\phi\right)\,\delta\left(
z-\min[z_b,1-z_b]
\right)\nonumber\\
&+\frac{4}{(4\pi)^3}\int  d\theta_{\bar b g}^2\left(
\frac{1}{\theta_{\bar b g}^2}
\right)_+\,d\theta^2_{b\bar b}\, d\phi\,dz_{\bar b}\,dz_g\left(
\frac{1}{z_g}
\right)_+\,z_b\,z_{\bar b}\,\frac{\theta_{b\bar b}^2}{\theta_{b\bar b}^2+\theta_{\bar b g}^2-2\theta_{b\bar b}\theta_{\bar b g}\cos\phi}\nonumber\\
&\hspace{1cm}
\times \delta(1-z_b-z_{\bar b}-z_g)\, \Theta\left(z_bz_{\bar b}\theta_{b\bar b}^2-\min[z_bz_g\theta_{bg}^2,z_{\bar b}z_g\theta_{\bar b g}^2]\right)\delta\left(
z-\min[z_b,1-z_b]
\right)\nonumber\\
&\hspace{1cm}\times\delta\left(
\frac{m_H^2}{E^2}-z_bz_{\bar b}\theta_{b\bar b}^2-z_{\bar b}z_g\theta_{\bar b g}^2-z_bz_g(\theta_{b\bar b}^2+\theta_{\bar b g}^2-2\theta_{b\bar b}\theta_{\bar b g}\cos\phi)
\right)\Theta\left(g_\text{alg}(z_b,z_g)\, z_b\theta_{bg}^2-g_\text{alg}(z_{\bar b},z_g)\,z_{\bar b}\theta_{\bar b g}^2\right)
\,.\nonumber
\end{align}
Here, we have combined the collinear divergent term (with the $\delta(\theta_{\bar b g}^2)$ factor) and the soft and collinear term (with both $\delta(\theta_{\bar b g}^2)$ and $\delta(z_g)$ factors) The final integral is explicitly finite, and so can be evaluated with standard Monte Carlo methods.  The second integral can simplified by rescaling the angle as
\begin{align}
\theta_{\bar b g}^2 \to \frac{m_H^2}{z_b(1-z_b)E^2}\, \theta_{\bar b g}^2\,.
\end{align}
Then, the integrals can be expressed as
\begin{align}
&\int \frac{d\Phi^{(1)}}{d\Phi^{(0)}}\, p^{(1,\text{sing})}(\Phi^{(1)})\\
&\supset   \frac{(m_H^2)^{-\epsilon}}{(4\pi)^{2-\epsilon}}\left(\frac{1}{\epsilon}+4-\frac{\pi^2}{3}-2\log(1-z_b)\right) \delta\left(
z-\min[z_b,1-z_b]
\right)\nonumber\\
&-\frac{4^{1-\epsilon}(m_H^2)^{-\epsilon}}{(4\pi)^{3-\epsilon}}\frac{\Gamma(1-\epsilon)}{\Gamma(1-2\epsilon)}\frac{1}{2\epsilon}\int  \frac{d\theta_{\bar b g}^2}{(\theta_{\bar b g}^2)^{1+\epsilon}}\, d\phi\,z_b^\epsilon\, (1-z_b)^{-\epsilon}\,\frac{\sin^{-2\epsilon}\phi}{1+\theta_{\bar b g}^2-2\theta_{\bar b g}\cos\phi}\nonumber\\
&\hspace{7cm}
\times \Theta\left(1-2\theta_{\bar b g}\cos\phi\right)\,\delta\left(
z-\min[z_b,1-z_b]
\right)\nonumber\\
&+\frac{4}{(4\pi)^3}\int  d\theta_{\bar b g}^2\left(
\frac{1}{\theta_{\bar b g}^2}
\right)_+\,d\theta^2_{b\bar b}\, d\phi\,dz_{\bar b}\,dz_g\left(
\frac{1}{z_g}
\right)_+\,z_b\,z_{\bar b}\,\frac{\theta_{b\bar b}^2}{\theta_{b\bar b}^2+\theta_{\bar b g}^2-2\theta_{b\bar b}\theta_{\bar b g}\cos\phi}\nonumber\\
&\hspace{1cm}
\times \delta(1-z_b-z_{\bar b}-z_g)\, \Theta\left(z_bz_{\bar b}\theta_{b\bar b}^2-\min[z_bz_g\theta_{bg}^2,z_{\bar b}z_g\theta_{\bar b g}^2]\right)\delta\left(
z-\min[z_b,1-z_b]
\right)\nonumber\\
&\hspace{1cm}\times\delta\left(
\frac{m_H^2}{E^2}-z_bz_{\bar b}\theta_{b\bar b}^2-z_{\bar b}z_g\theta_{\bar b g}^2-z_bz_g(\theta_{b\bar b}^2+\theta_{\bar b g}^2-2\theta_{b\bar b}\theta_{\bar b g}\cos\phi)
\right)\Theta\left(g_\text{alg}(z_b,z_g)\, z_b\theta_{bg}^2-g_\text{alg}(z_{\bar b},z_g)\,z_{\bar b}\theta_{\bar b g}^2\right)
\,.\nonumber
\end{align}

The remaining integral in which divergences must be extracted is
\begin{align}
\int  \frac{d\theta_{\bar b g}^2}{(\theta_{\bar b g}^2)^{1+\epsilon}}\, d\phi\,z_b^\epsilon\, (1-z_b)^{-\epsilon}\,\frac{\sin^{-2\epsilon}\phi}{1+\theta_{\bar b g}^2-2\theta_{\bar b g}\cos\phi}\, \Theta\left(1-2\theta_{\bar b g}\cos\phi\right)\,,
\end{align}
which has a non-trivial angular constraint.  Note that
\begin{align}
\Theta\left(1-2\theta_{\bar b g}\cos\phi\right)&=\Theta\left(1-2\theta_{\bar b g}\right)+\Theta\left(2\theta_{\bar b g}-1\right)\Theta\left(1-2\theta_{\bar b g}\cos\phi\right)
\end{align}
The second constraint eliminates collinear divergences, so we can write
\begin{align}
&\int  \frac{d\theta_{\bar b g}^2}{(\theta_{\bar b g}^2)^{1+\epsilon}}\, d\phi\,z_b^\epsilon\, (1-z_b)^{-\epsilon}\,\frac{\sin^{-2\epsilon}\phi}{1+\theta_{\bar b g}^2-2\theta_{\bar b g}\cos\phi}\, \Theta\left(1-2\theta_{\bar b g}\cos\phi\right)\\
&\hspace{1cm}=\int  \frac{d\theta_{\bar b g}^2}{(\theta_{\bar b g}^2)^{1+\epsilon}}\, d\phi\,z_b^\epsilon\, (1-z_b)^{-\epsilon}\,\frac{\sin^{-2\epsilon}\phi}{1+\theta_{\bar b g}^2-2\theta_{\bar b g}\cos\phi}\, \Theta\left(1-2\theta_{\bar b g}\right)\nonumber\\
&\hspace{2cm}
+\int  \frac{d\theta_{\bar b g}^2}{\theta_{\bar b g}^2}\, d\phi\,z_b^\epsilon\, (1-z_b)^{-\epsilon}\,\frac{\sin^{-2\epsilon}\phi}{1+\theta_{\bar b g}^2-2\theta_{\bar b g}\cos\phi}\,\Theta\left(2\theta_{\bar b g}-1\right)\Theta\left(1-2\theta_{\bar b g}\cos\phi\right)\nonumber\,.
\end{align}
Additionally, note that the expansion in $\epsilon$ of the $z_b$-dependent term is
\begin{align}
z_b^\epsilon\, (1-z_b)^{-\epsilon} = 1+\epsilon\log\frac{z_b}{1-z_b}+\frac{\epsilon^2}{2}\log^2\frac{z_b}{1-z_b}+\cdots\,.
\end{align}
Then, in the NLO-to-LO map, we define the energy fraction $z$ as the smallest of $z_b$ and $1-z_b$, which renders the resulting expression symmetric in $z\leftrightarrow 1-z$.  Therefore, any term that is anti-symmetric in $z_b\leftrightarrow 1-z_b$ does not survive the NLO-to-LO map, and can be safely discarded.  Thus, there is no term linear in $\epsilon$ in this expansion.  The elimination of such a term is critical for establishing IRC safety because divergent terms can only be safely discarded if they are independent of $z_b$ which then ensures that real and virtual divergences perfectly cancel.  

Using this observation, the $\epsilon$ expansion of this integral is
\begin{align}
&\int  \frac{d\theta_{\bar b g}^2}{(\theta_{\bar b g}^2)^{1+\epsilon}}\, d\phi\,z_b^\epsilon\, (1-z_b)^{-\epsilon}\,\frac{\sin^{-2\epsilon}\phi}{1+\theta_{\bar b g}^2-2\theta_{\bar b g}\cos\phi}\, \Theta\left(1-2\theta_{\bar b g}\cos\phi\right)\\
&\hspace{1cm}=\int  \frac{d\theta_{\bar b g}^2}{(\theta_{\bar b g}^2)^{1+\epsilon}}\, d\phi\left(1+\frac{\epsilon^2}{2}\log^2\frac{z_b}{1-z_b}\right)\frac{\sin^{-2\epsilon}\phi}{1+\theta_{\bar b g}^2-2\theta_{\bar b g}\cos\phi}\, \Theta\left(1-2\theta_{\bar b g}\right)\nonumber\\
&\hspace{2cm}
+\int  \frac{d\theta_{\bar b g}^2}{\theta_{\bar b g}^2}\, d\phi\,\frac{1-2\epsilon\log\left(\sin\phi\right)}{1+\theta_{\bar b g}^2-2\theta_{\bar b g}\cos\phi}\,\Theta\left(2\theta_{\bar b g}-1\right)\Theta\left(1-2\theta_{\bar b g}\cos\phi\right)\nonumber\,.
\end{align}
To evaluate the first integral as an expansion in $\epsilon$, we introduce the $+$-function expansion, where, for convenience, we define
\begin{align}
\frac{1}{(\theta_{\bar b g}^2)^{1+\epsilon}} = -\frac{4^\epsilon}{\epsilon}\,\delta(\theta_{\bar bg}^2)+\left(
\frac{1}{\theta_{\bar bg}^2}
\right)_+-\epsilon\left(\frac{\log\theta_{\bar b g}^2}{\theta_{\bar b g}^2}\right)_++\cdots\,,
\end{align}
where the $+$-functions integrate to 0 on $\theta_{\bar b g}^2\in[0,1/4]$.  This choice makes the angular constraint as simple as possible.  Further, as discussed above, we only need to keep those terms that explicitly depend on $z_b$ in a way that is $z_b\leftrightarrow 1-z_b$ symmetric, and so we can ignore many constant terms.  The integral with these simplifications then dramatically reduces to
\begin{align}
&\int  \frac{d\theta_{\bar b g}^2}{(\theta_{\bar b g}^2)^{1+\epsilon}}\, d\phi\,z_b^\epsilon\, (1-z_b)^{-\epsilon}\,\frac{\sin^{-2\epsilon}\phi}{1+\theta_{\bar b g}^2-2\theta_{\bar b g}\cos\phi}\, \Theta\left(1-2\theta_{\bar b g}\cos\phi\right)\supset-\epsilon\frac{\pi}{2}\log^2\frac{z_b}{1-z_b}\,.
\end{align}

Now, putting all of these results together and explicitly eliminating terms independent of $z_b$, we find
\begin{align}
&\int \frac{d\Phi^{(1)}}{d\Phi^{(0)}}\, p^{(1,\text{sing})}(\Phi^{(1)})\\
&\supset   -\frac{2}{(4\pi)^2}\log(1-z_b)\, \delta\left(
z-\min[z_b,1-z_b]
\right)+\frac{1}{4(4\pi)^2}\log^2\frac{z_b}{1-z_b}\,\delta\left(
z-\min[z_b,1-z_b]
\right)\nonumber\\
&+\frac{4}{(4\pi)^3}\int  d\theta_{\bar b g}^2\left(
\frac{1}{\theta_{\bar b g}^2}
\right)_+\,d\theta^2_{b\bar b}\, d\phi\,dz_{\bar b}\,dz_g\left(
\frac{1}{z_g}
\right)_+\,z_b\,z_{\bar b}\,\frac{\theta_{b\bar b}^2}{\theta_{b\bar b}^2+\theta_{\bar b g}^2-2\theta_{b\bar b}\theta_{\bar b g}\cos\phi}\nonumber\\
&\hspace{1cm}
\times \delta(1-z_b-z_{\bar b}-z_g)\, \Theta\left(z_bz_{\bar b}\theta_{b\bar b}^2-\min[z_bz_g\theta_{bg}^2,z_{\bar b}z_g\theta_{\bar b g}^2]\right)\delta\left(
z-\min[z_b,1-z_b]
\right)\nonumber\\
&\hspace{1cm}\times\delta\left(
\frac{m_H^2}{E^2}-z_bz_{\bar b}\theta_{b\bar b}^2-z_{\bar b}z_g\theta_{\bar b g}^2-z_bz_g(\theta_{b\bar b}^2+\theta_{\bar b g}^2-2\theta_{b\bar b}\theta_{\bar b g}\cos\phi)
\right)\Theta\left(g_\text{alg}(z_b,z_g)\, z_b\theta_{bg}^2-g_\text{alg}(z_{\bar b},z_g)\,z_{\bar b}\theta_{\bar b g}^2\right)
\,.\nonumber
\end{align}
This is manifestly finite, explicitly demonstrating IRC safety.  The contribution to the cumulative distribution is thus
\begin{align}
&I_3^{(1,\text{sing})}({\cal L}) = \frac{4}{(4\pi)^2}\,C_F\int_0^{1/2} dz\, p^{(0)}(z) \left[\Theta\left(
{\cal L}-\hat {\cal L}^{(0)}(z)
\right)-\Sigma^{(0)}({\cal L})\right]\\
&\times \Biggl[
-2\log z-2\log(1-z)+\frac{1}{2}\log^2\frac{z}{1-z}\nonumber\\
&
+\frac{1}{\pi}\int  d\theta_{\bar b g}^2\left(
\frac{1}{\theta_{\bar b g}^2}
\right)_+\,d\theta^2_{b\bar b}\, d\phi\,dz_b\,dz_{\bar b}\,dz_g\left(
\frac{1}{z_g}
\right)_+\,z_b\,z_{\bar b}\,\frac{\theta_{b\bar b}^2}{\theta_{b\bar b}^2+\theta_{\bar b g}^2-2\theta_{b\bar b}\theta_{\bar b g}\cos\phi}\nonumber\\
&\hspace{1cm}
\times \delta(1-z_b-z_{\bar b}-z_g)\, \Theta\left(z_bz_{\bar b}\theta_{b\bar b}^2-\min[z_bz_g\theta_{bg}^2,z_{\bar b}z_g\theta_{\bar b g}^2]\right)\delta\left(
z-\min[z_b,1-z_b]
\right)\nonumber\\
&\hspace{1cm}\times\delta\left(
\frac{m_H^2}{E^2}-z_bz_{\bar b}\theta_{b\bar b}^2-z_{\bar b}z_g\theta_{\bar b g}^2-z_bz_g(\theta_{b\bar b}^2+\theta_{\bar b g}^2-2\theta_{b\bar b}\theta_{\bar b g}\cos\phi)
\right)\Theta\left(g_\text{alg}(z_b,z_g)\, z_b\theta_{bg}^2-g_\text{alg}(z_{\bar b},z_g)\,z_{\bar b}\theta_{\bar b g}^2\right)
\Biggr]\nonumber\,.
\end{align}

\subsection{Non-Abelian Soft Term}

The non-Abelian soft term, that only appears because the initial gluon is a color octet, is
\begin{align}
I_4^{(1,\text{sing})}({\cal L}) &\equiv 2C_A\int_0^{1/2} dz\, p^{(0)}(z) \left[\Theta\left(
{\cal L}-\hat {\cal L}^{(0)}(z)
\right)-\Sigma^{(0)}({\cal L})\right] \int \frac{d\Phi^{(1)}}{d\Phi^{(0)}}\, \frac{z_bs_{\bar b g}+z_{\bar b}s_{b g}-z_g s_{b\bar b}}{z_gs_{bg}s_{\bar b g}}\nonumber\\
&\hspace{-1.5cm}\times\Theta\left(s_{b\bar b}-\min[s_{b g},s_{\bar b g}]\right)\Theta\left(g_\text{alg}(z_b,z_g)s_{bg}-g_\text{alg}(z_{\bar b},z_g)\, s_{\bar b g}\right)\delta\left(
z-\min[z_b,1-z_b]
\right)\,.
\end{align}
This term has a soft divergence, $z_g\to 0$, that must be regulated.  In angle-energy fraction coordinates, the integral over phase space of this term is
\begin{align}
&\int \frac{d\Phi^{(1)}}{d\Phi^{(0)}}\, p^{(1,\text{sing})}(\Phi^{(1)})\\
&\supset  \frac{4^{1-\epsilon}(m_H^2)^\epsilon(E^2)^{-2\epsilon}}{(4\pi)^{3-\epsilon}}\frac{\Gamma(1-\epsilon)}{\Gamma(1-2\epsilon)}\int  d\theta_{\bar b g}^2\,d\theta^2_{b\bar b}\, d\phi\,dz_{\bar b}\,\frac{dz_g}{z_g^{1+2\epsilon}}\,z_b^{1-\epsilon}\,(1-z_b)^\epsilon\,z_{\bar b}^{1-2\epsilon}\nonumber\\
&\hspace{1cm}\times\frac{(\theta_{b\bar b}^2)^{-\epsilon}(\theta_{\bar b g}^2)^{-1-\epsilon}\sin^{-2\epsilon}\phi}{\theta_{b\bar b}^2+\theta_{\bar b g}^2-2\theta_{b\bar b}\theta_{\bar b g}\cos\phi}\left(
2\theta_{\bar b g}^2-2\theta_{b\bar b}\theta_{\bar bg}\cos\phi
\right)\nonumber\\
&\hspace{1cm}
\times\delta\left(
\frac{m_H^2}{E^2}-z_bz_{\bar b}\theta_{b\bar b}^2-z_{\bar b}z_g\theta_{\bar b g}^2-z_bz_g(\theta_{b\bar b}^2+\theta_{\bar b g}^2-2\theta_{b\bar b}\theta_{\bar b g}\cos\phi)
\right)\delta(1-z_b-z_{\bar b}-z_g)\nonumber\\
&\hspace{1cm}\times \Theta\left(z_bz_{\bar b}\theta_{b\bar b}^2-\min[z_bz_g\theta_{bg}^2,z_{\bar b}z_g\theta_{\bar b g}^2]\right)\Theta\left(g_\text{alg}(z_b,z_g)\, z_b\theta_{bg}^2-g_\text{alg}(z_{\bar b},z_g)\,z_{\bar b}\theta_{\bar b g}^2\right)\nonumber\\
&=\frac{4^{1-\epsilon}(m_H^2)^\epsilon(E^2)^{-2\epsilon}}{(4\pi)^{3-\epsilon}}\frac{\Gamma(1-\epsilon)}{\Gamma(1-2\epsilon)}\int  d\theta_{\bar b g}^2\,d\theta^2_{b\bar b}\, d\phi\,dz_{\bar b}\,dz_g\left(
-\frac{(1-z_b)^{-2\epsilon}}{2\epsilon}\delta(z_g)
\right)\,z_b^{1-\epsilon}\,(1-z_b)^\epsilon\,z_{\bar b}^{1-2\epsilon}\nonumber\\
&\hspace{1cm}\times\frac{(\theta_{b\bar b}^2)^{-\epsilon}(\theta_{\bar b g}^2)^{-1-\epsilon}\sin^{-2\epsilon}\phi}{\theta_{b\bar b}^2+\theta_{\bar b g}^2-2\theta_{b\bar b}\theta_{\bar b g}\cos\phi}\left(
2\theta_{\bar b g}^2-2\theta_{b\bar b}\theta_{\bar bg}\cos\phi
\right)\nonumber\\
&\hspace{1cm}
\times\delta\left(
\frac{m_H^2}{E^2}-z_bz_{\bar b}\theta_{b\bar b}^2-z_{\bar b}z_g\theta_{\bar b g}^2-z_bz_g(\theta_{b\bar b}^2+\theta_{\bar b g}^2-2\theta_{b\bar b}\theta_{\bar b g}\cos\phi)
\right)\delta(1-z_b-z_{\bar b}-z_g)\nonumber\\
&\hspace{1cm}\times \Theta\left(z_bz_{\bar b}\theta_{b\bar b}^2-\min[z_bz_g\theta_{bg}^2,z_{\bar b}z_g\theta_{\bar b g}^2]\right)\Theta\left(g_\text{alg}(z_b,z_g)\, z_b\theta_{bg}^2-g_\text{alg}(z_{\bar b},z_g)\,z_{\bar b}\theta_{\bar b g}^2\right)\nonumber\\
&+\frac{4}{(4\pi)^3}\int  \frac{d\theta_{\bar b g}^2}{\theta_{\bar b g}^2}\,d\theta^2_{b\bar b}\, d\phi\,dz_{\bar b}\,dz_g\left(
\frac{1}{z_g}
\right)_+\,z_b\,z_{\bar b}\,\frac{2\theta_{\bar b g}^2-2\theta_{b\bar b}\theta_{\bar bg}\cos\phi}{\theta_{b\bar b}^2+\theta_{\bar b g}^2-2\theta_{b\bar b}\theta_{\bar b g}\cos\phi}\nonumber\\
&\hspace{1cm}
\times\delta\left(
\frac{m_H^2}{E^2}-z_bz_{\bar b}\theta_{b\bar b}^2-z_{\bar b}z_g\theta_{\bar b g}^2-z_bz_g(\theta_{b\bar b}^2+\theta_{\bar b g}^2-2\theta_{b\bar b}\theta_{\bar b g}\cos\phi)
\right)\delta(1-z_b-z_{\bar b}-z_g)\nonumber\\
&\hspace{1cm}\times \Theta\left(z_bz_{\bar b}\theta_{b\bar b}^2-\min[z_bz_g\theta_{bg}^2,z_{\bar b}z_g\theta_{\bar b g}^2]\right)\Theta\left(g_\text{alg}(z_b,z_g)\, z_b\theta_{bg}^2-g_\text{alg}(z_{\bar b},z_g)\,z_{\bar b}\theta_{\bar b g}^2\right)
\nonumber\,.
\end{align}
Isolating the divergent term, we have
\begin{align}
&\int \frac{d\Phi^{(1)}}{d\Phi^{(0)}}\, p^{(1,\text{sing})}(\Phi^{(1)})\\
&\supset -\frac{4^{1-\epsilon}(m_H^2)^\epsilon(E^2)^{-2\epsilon}}{(4\pi)^{3-\epsilon}}\frac{1}{2\epsilon}\int  d\theta_{\bar b g}^2\,d\theta^2_{b\bar b}\, d\phi\,dz_{\bar b}
\,z_b^{1-\epsilon}\,(1-z_b)^{1-3\epsilon}\nonumber\\
&\hspace{1cm}\times\frac{(\theta_{b\bar b}^2)^{-\epsilon}(\theta_{\bar b g}^2)^{-1-\epsilon}\sin^{-2\epsilon}\phi}{\theta_{b\bar b}^2+\theta_{\bar b g}^2-2\theta_{b\bar b}\theta_{\bar b g}\cos\phi}\left(
2\theta_{\bar b g}^2-2\theta_{b\bar b}\theta_{\bar bg}\cos\phi
\right)\delta\left(
\frac{m_H^2}{E^2}-z_b(1-z_b)\theta_{b\bar b}^2
\right)\nonumber\\
&\hspace{1cm}\times \Theta\left(\theta_{b\bar b}-2\theta_{\bar b g}\cos\phi\right)\nonumber
\nonumber\,.
\end{align}
To separate angle and energy factors, we rescale both angles as
\begin{align}
\theta^2 \to \frac{m_H^2}{z_b(1-z_b)E^2}\, \theta^2\,.
\end{align}
The divergent term then reduces to
\begin{align}
&\int \frac{d\Phi^{(1)}}{d\Phi^{(0)}}\, p^{(1,\text{sing})}(\Phi^{(1)})\supset -\frac{4^{1-\epsilon}(m_H^2)^{-\epsilon}}{(4\pi)^{3-\epsilon}}\frac{1}{2\epsilon}\int  d\theta_{\bar b g}^2\,d\theta^2_{b\bar b}\, d\phi\,dz_{\bar b}
\,z_b^{\epsilon}\,(1-z_b)^{-\epsilon}\\
&\hspace{1cm}\times\frac{(\theta_{b\bar b}^2)^{-\epsilon}(\theta_{\bar b g}^2)^{-1-\epsilon}\sin^{-2\epsilon}\phi}{\theta_{b\bar b}^2+\theta_{\bar b g}^2-2\theta_{b\bar b}\theta_{\bar b g}\cos\phi}\left(
2\theta_{\bar b g}^2-2\theta_{b\bar b}\theta_{\bar bg}\cos\phi
\right)\delta\left(
1-\theta_{b\bar b}^2
\right) \Theta\left(\theta_{b\bar b}-2\theta_{\bar b g}\cos\phi\right)\nonumber
\nonumber\,.
\end{align}
As discussed in the previous section, the $\epsilon$ expansion of the factor $z_b^{\epsilon}\,(1-z_b)^{-\epsilon}$ has no term at order-$\epsilon$ that contributes to the cumulative distribution of the likelihood.  As such, this entire term is actually independent of $z_b$, and so can be safely ignored.  Therefore, the only contribution to this integral is
\begin{align}
&I_4^{(1,\text{sing})}({\cal L}) = \frac{2}{(4\pi)^2}\,C_A\int_0^{1/2} dz\, p^{(0)}(z) \left[\Theta\left(
{\cal L}-\hat {\cal L}^{(0)}(z)
\right)-\Sigma^{(0)}({\cal L})\right]\\
&\times \frac{1}{\pi}\int  \frac{d\theta_{\bar b g}^2}{\theta_{\bar b g}^2}\,d\theta^2_{b\bar b}\, d\phi\,dz_b\,dz_{\bar b}\,dz_g\left(
\frac{1}{z_g}
\right)_+\,z_b\,z_{\bar b}\,\frac{2\theta_{\bar b g}^2-2\theta_{b\bar b}\theta_{\bar bg}\cos\phi}{\theta_{b\bar b}^2+\theta_{\bar b g}^2-2\theta_{b\bar b}\theta_{\bar b g}\cos\phi}\,\delta(z-\min[z_b,1-z_b])\nonumber\\
&\hspace{1cm}
\times\delta\left(
\frac{m_H^2}{E^2}-z_bz_{\bar b}\theta_{b\bar b}^2-z_{\bar b}z_g\theta_{\bar b g}^2-z_bz_g(\theta_{b\bar b}^2+\theta_{\bar b g}^2-2\theta_{b\bar b}\theta_{\bar b g}\cos\phi)
\right)\delta(1-z_b-z_{\bar b}-z_g)\nonumber\\
&\hspace{1cm}\times \Theta\left(z_bz_{\bar b}\theta_{b\bar b}^2-\min[z_bz_g\theta_{bg}^2,z_{\bar b}z_g\theta_{\bar b g}^2]\right)\Theta\left(g_\text{alg}(z_b,z_g)\, z_b\theta_{bg}^2-g_\text{alg}(z_{\bar b},z_g)\,z_{\bar b}\theta_{\bar b g}^2\right)\nonumber\,.
\end{align}

\bibliography{nlodisc}

\end{document}